\documentclass{nature}

\usepackage{times}
\usepackage{caption}
\usepackage{graphicx}
\usepackage{amsmath}
\usepackage{amssymb}
\usepackage[dvipsnames]{xcolor}
\usepackage{ulem}
\usepackage{eucal}
\usepackage{lineno}
\usepackage{float}
\usepackage{changes}
\usepackage{xcolor}%

\usepackage{dcolumn}
\usepackage{bm}

\title{
Elastocapillary sequential fluid capture in hummingbird-inspired grooved sheets} 
\author
{Emmanuel Si\'{e}fert$^{1}$, Benoit Scheid$^{2}$, Fabian Brau$^{1}$, Jean Cappello$^{2,3}$}

\begin{document} 
\baselineskip24pt
\maketitle 
\begin{affiliations}
\item Université libre de Bruxelles (ULB), Nonlinear Physical Chemistry Unit, CP231, 1050 Bruxelles, Belgium
\item Université libre de Bruxelles (ULB), TIPS, CP165/67, 1050, Bruxelles, Belgium
\item Universit\'e Claude Bernard Lyon 1, CNRS, Institut Lumi\`ere Mati\`ere, F-69622 Villeurbanne, France.
\end{affiliations}


\begin{abstract}

Passive and effective fluid capture and transport at small scale is crucial for industrial and medical applications, especially for the realisation of  point-of-care tests. 
Performing these tests involves several steps including biological fluid capture, aliquoting, reaction with reagents at the fluid-device interface, and reading of the results. Ideally, these tests must be fast and offer a large surface-to-volume ratio to achieve rapid and precise diagnostics with a reduced amount of fluid. Such constraints are often contradictory as a high surface-to-volume ratio implies a high hydraulic resistance and hence a decrease in the flow rate. Inspired by the feeding mechanism of hummingbirds, we propose a frugal fluid capture device that takes advantage of elastocapillary deformations to enable concomitant fast liquid transport, aliquoting, and high confinement in the deformed state. The hierarchical design of the device -- that consists in vertical grooves stacked on an elastic sheet -- enables a two-step sequential fluid capture. Each unit groove mimics the hummingbird's tongue and closes due to capillary forces when a wetting liquid penetrates, yielding the closure of the whole device in a tubular shape, where additional liquid is captured. Combining elasticity, capillarity, and viscous flow, we rationalise the fluid-structure interaction at play both when liquid is scarce -- end dipping and capillary rise -- and abundant -- full dipping. By functionalising the surface of the grooves such a passive device can concomitantly achieve all the steps of point-of-care tests, opening the way for the design of optimal devices for fluid capture and transport in microfluidics.

\end{abstract}
When a porous medium is put in contact with a wetting liquid, capillary forces lead to the imbibition of the liquid inside the pores of the material, overcoming gravity\cite{lago2001capillary}. This phenomenon of capillary rise is ubiquitous: it brings water to the upper layer of soils\cite{lu2004rate}, drives sap in plants\cite{jensen2016sap}, and is crucial in textile and paper industries\cite{alava2006physics,duprat2022moisture}. 
Fundamental capillarity laws were obtained a century ago by Lucas\cite{lucas1918ueber} and Washburn\cite{washburn1921dynamics}, by balancing the driving capillary force and viscous friction in horizontal pores. 
This basic modelling approach has been further enriched in a vast literature, to model the capillary rise in pipes of various orientations with gravitational effects\cite{zhmud2000dynamics, Fries2008} and various closed\cite{Ouali2013, Berthier2015} or open\cite{Mann1995,yang2011dynamics,kolliopoulos2021capillary,kim2022evaporative} 
geometries, 
 to regularise the early dynamics via inertial\cite{quere1997inertial, Quere1999} 
 or contact line friction\cite{delannoy2019dual} effects,
and to model the effects of non-Newtonian liquids\cite{Berli2014, Yang2020}. 
Recently, bioinspired asymmetric textures were designed to obtain directional capillary transport along specific directions\cite{feng2021three,li2019bioinspired}. However, the imbibing speed is set by the typical pore size, yielding slow liquid penetration at small scales, that are particularly relevant for medical testing devices since they offer a large surface-to-volume ratio\cite{jung2015point}. To resolve these conflicting physical constraints and obtain fast penetration in confined pores, one possible approach relies on passively changing the pore shape and size using a soft material, such that the pores are relatively larger in their undeformed state, promoting fast liquid imbibition, and smaller in their deformed state, offering a confined environment beneficial for testing purposes. Soft porous solids, that may be deformed by capillary forces leading to a coupling between the deformable porous medium and the liquid\cite{bico2018elastocapillarity,bico2004elastocapillary,li2021liquid}, are ideal candidates to meet these needs.
Yet, so far, studies mostly focused on the coalescence of the pores, leading to an enhanced but slower capillary rise\cite{cambau2011capillary,duprat2011dynamics}.
 
\begin{figure}[!ht]
  \centering
    \includegraphics[width=0.8\textwidth]{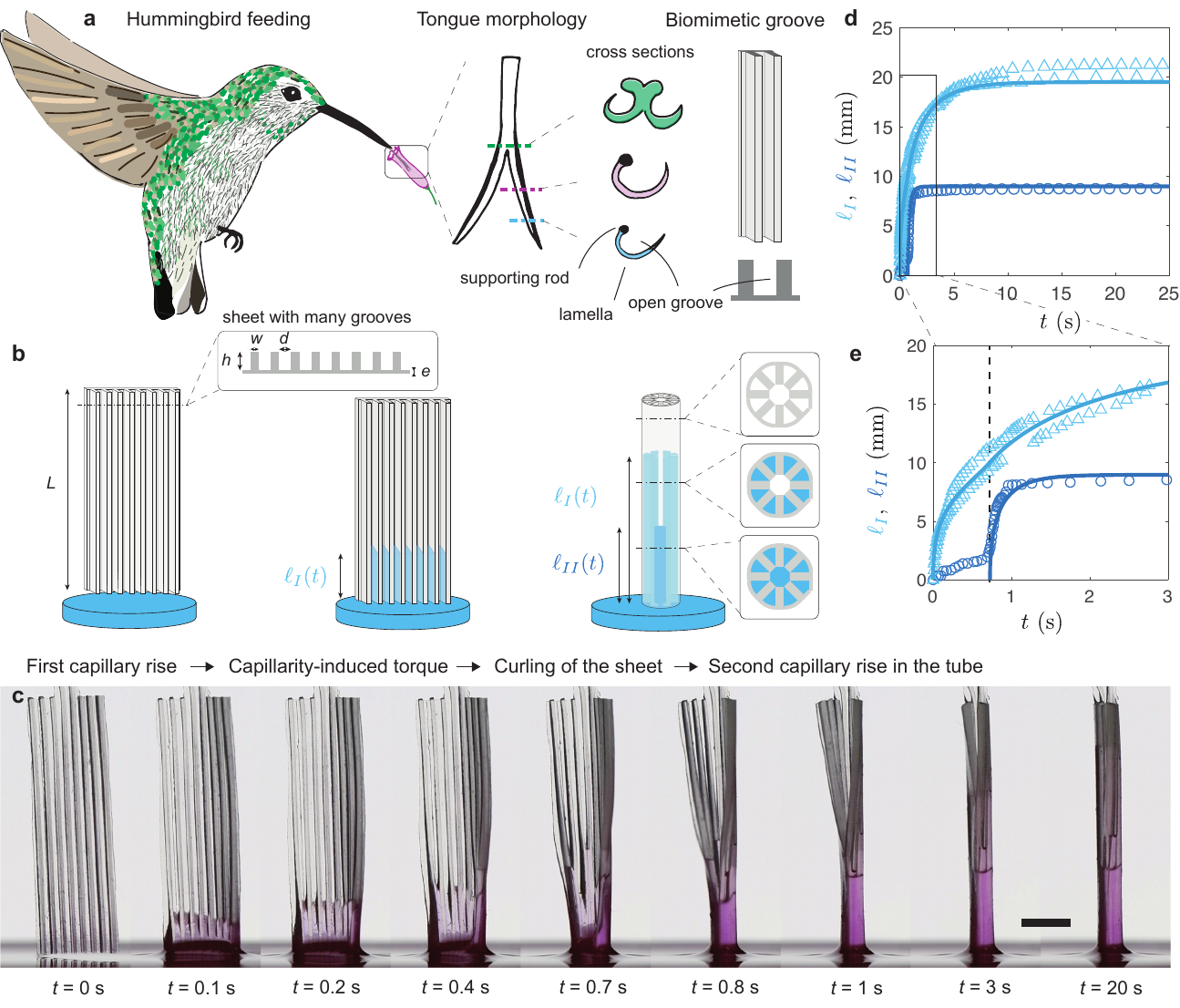}
      \caption{ {\bf Hummingbird-inspired sequential elastocapillary rise.} {\bf a}, Hummingbirds feed by dipping into nectar their flexible tongue made of a pair of pre-curved lamellae forming open grooves. {\bf b}, Biomimetic groovy sheet of length $L$, thickness $e$, decorated with walls of height $h$, width $w$, forming grooves of width $d$, as it is dipped into a wetting fluid bath (left). A first capillary rise (characterised by the front elevation $\ell_I(t)$) occurs inside the grooves (center), inducing a capillary torque on the sheet that bends into a tubular shape (right). A second capillary rise ($\ell_{II}(t)$) takes then place in the newly formed tube. {\bf c}, Snapshots of the hierarchical capillary rise in a grooved sheet ($e=135\pm10$ µm, $h=800\pm50$ µm, $w=400\pm40$ µm, $d=700\pm40$ µm, $L=26.1\pm 0.2$ mm) made of a silicone rubber. Liquid used is silicone oil V10 ($\mu=9.5 \times 10^{-3}$ Pa s, $\gamma=0.021$ N/m). Scale bar: 5~mm (see Supplementary Video 1 and 2). {\bf d}, Capillary rise as a function of time in two different grooves of the same textured sheet ($\ell_I$, triangles) and in the tube formed when the structure is closed ($\ell_{II}$, circles). The solid lines correspond to the model without any fitting parameter for $e=129$ µm, $h=800$ µm, $w=430$ µm, $d=670$ µm, $L=26$ mm. {\bf e}, Zoom on the early dynamics, with the closing time $t_\text{closure}$ highlighted by a vertical dashed line.
      }
      \label{fig1}
\end{figure}

Hummingbird tongues offer an inspiring natural example of a deformable structure that meets the needs of fast liquid capture.
To feed, hummingbirds need to hover above flowers~\cite{warrick2005aerodynamics}, the most energy consuming form of locomotion in the animal kingdom\cite{weis1972energetics}. The feeding strategy of this small nectarivore thus has to be fast and efficient. Evolution led to the unique shape of the hummingbird tongue: it is composed of two flexible curved lamellae fixed to a supporting rod and forming an open groove (Fig.~\ref{fig1}a). In vivo observations reveal that these grooves act as dynamic elastocapillary liquid-trapping devices\cite{rico2011hummingbird} that passively close into tubes as it is withdrawn from an abundant nectar source. When nectar is scarce however, the tongue efficiently acts as a capillary syphon\cite{kim11,kim2012hummingbird}, leading to a fast capillary rise in the closing grooves that may be actively enhanced by a micropump effect\cite{rico2015hummingbird}.

Inspired by hummingbirds, we go one step further and design flat architected elastic sheets (Fig.~\ref{fig1}a,b) that contain many vertical grooves, each of them acting as the curved lamella of the hummingbird's tongue. This ``metatongue" takes advantage of the collective elastocapillary deformation of each groove --that plays the role of an open pore-- to deform at the device scale. Details on the fabrication of these structures are given in Methods section `Elastomer preparation'. Depending on the amount of liquid available, the device is either slightly put in contact or fully dipped into the liquid (Methods section `Experimental apparatus'). When liquid is scarce, as contact is made with the liquid, a first capillary rise occurs in the grooves (first instants in Fig.~\ref{fig1}b and c). This first capillary rise of height $\ell_I$ (light blue curves in Fig.~\ref{fig1}d) induces a capillary torque on the bottom sheet of each groove and hence its bending. This bending brings adjacent walls closer and closes each groove into a pipe of circular sector cross-section, trapping the liquid as in the case of the hummingbird tongue. As each groove deforms, the whole sheet consequently bends and may eventually close into a tubular shape, leading to a second capillary rise in the newly formed tube (final stages in Fig.~\ref{fig1}b and c, Supplementary Video 1 and 2). The dynamics (dark blue curves in Fig.~\ref{fig1}d) and final height of this second capillary rise  are intrinsically related to the groove geometry as the radius of the newly formed tube is given by the curling of the whole structure that stops when self contact between the neighbouring walls occurs\cite{cappello2023bioinspired}. When the liquid is abundant, a second liquid capture strategy consists in fully dipping the grooved sheet in the bath before withdrawing it (Fig.~\ref{fig3}). As the structure is  immersed, liquid enters the grooves from the side and the structure gets entirely imbibed. When removing the structure from the bath, capillary forces lead to its closing into a tube and hence a liquid capture in the grooves and in the newly formed tube. During the extrusion phase, the liquid drains from the pipes and so the withdrawing velocity plays a crucial role in the amount of captured liquid.

\begin{figure}[!ht]
  \centering
    \includegraphics[width=1\textwidth]{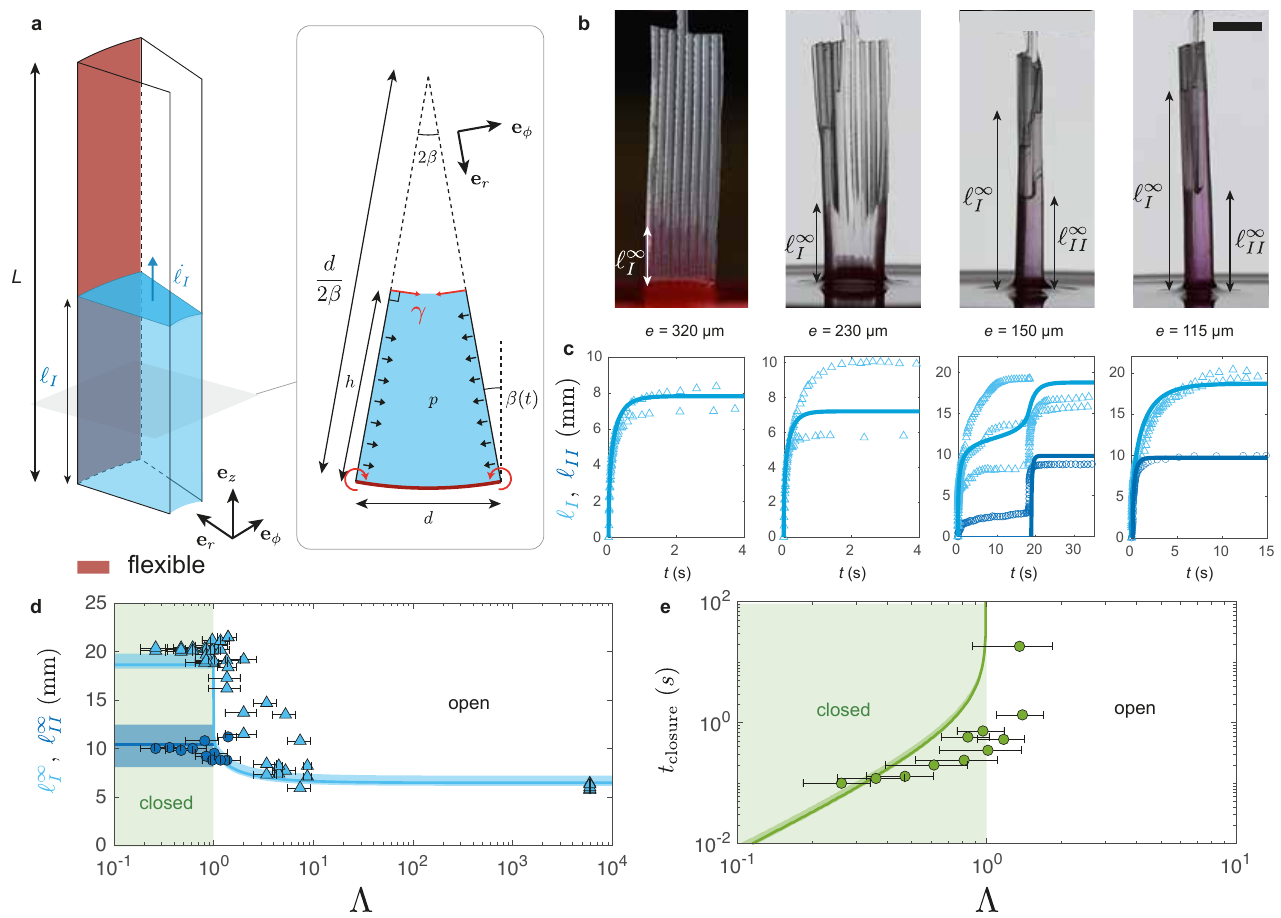}
\caption{ {\bf Modelling the fluid-structure interaction.} {\bf a}, Schematics of the flexible groove model: the capillary rise $\ell_I$ in the groove induces a capillary torque acting on the bottom flexible sheet of length $L$ (in red) via the walls considered as rigid, inducing a curvature radius $R=d/(2\beta)$. {\bf b}, Snapshot of the final equilibrium shape of the ribbed structure for various sheet thicknesses $e$ with the same rib geometry: $h=810\pm40$ µm, $w=400\pm40$ µm, $d=700\pm50$ µm. A transition to closure is observed for sufficiently thin sheets. {\bf c}, Corresponding evolution of the capillary rises $\ell_{I}$ and $\ell_{II}$ in two to three different grooves (triangles) and in the tube (circle) respectively. The solid lines correspond to the theory. {\bf d}, Final equilibrium elevations $\ell_I^{\, \infty}$ and $\ell_{II}^{\, \infty}$ as a function of the dimensionless transition parameter $\Lambda$ defined by Eq.~(\ref{lambda-def}). For both capillary rises, a sharp transition occurs at $\Lambda \sim 1$. {\bf e}, Time needed for the structure to close after contact with the liquid bath, $t_\text{closure}$, as a function of the transition parameter. This time diverges at the transition. Solid lines and shaded area correspond to the theory with the uncertainty on the geometrical parameters.}
       \label{fig2}
\end{figure}

We start by considering the situation where the liquid is scarce and the grooved sheet is slightly dipped in the liquid bath. Depending on the geometry of the grooved sheet (groove height $h$ and width $d$, wall width $w$ and sheet thickness $e$)  and on its mechanical properties (silicon rubber of Young's modulus $E$ and Poisson ratio $\nu$) three main scenarii can be observed while capillary rise occurs: (i) the structure remains undeformed and the capillary rise takes place in undeformed grooves of rectangular cross-section (first panel of Fig.~\ref{fig2}b), (ii) the sheet slightly bends due to capillary forces and the capillary rise occurs in an open channel as sketch in Fig.~\ref{fig2}a (second panel of Fig.~\ref{fig2}b), (iii) the sheet bends until neighbouring walls are in contact leading to the closure of the structure into a tube (last two panels in Fig.~\ref{fig2}b).
In the limit of a thick sheet (large $e$) or large Young modulus, i.e., when the sheet deformation is negligible (case (i)), the system is equivalent to the capillary rise inside rigid open grooves (Fig. \ref{fig2}a, with $\beta=0$) of depth $h$ and width $d$, that have been at the core of recent studies\cite{yang2011dynamics,kolliopoulos2021capillary,kim2022evaporative}. The equilibrium height $\ell^{\, \infty,\text{open}}_I$ is shown to barely depend on the depth $h$ of the channel as long as there is total wetting (i.e. $\theta_Y=0$ with $\theta_Y$ being the contact angle). Under these conditions, the equilibrium height reads
\begin{equation}
\ell^{\infty,\text{open}}_I=\frac{2\ell_c^2}{d},
    \label{eq:lopen}
\end{equation}
where $\ell_c=\sqrt{\gamma/(\rho g)}$ is the capillary length with $\gamma$ being the surface tension of the liquid, $\rho$ its density and $g$ the gravitational acceleration. The derivation of $\ell^{\, \infty,\text{open}}_I$ is given in the Methods section `Fluid-structure interaction and dimensionless equations'. Note that this elevation is equivalent to the equilibrium capillary rise between two infinite parallel plates separated by a distance $d$.
As previously mentioned, such a capillary rise induces a torque on the sheet through the walls. 
This torque has two contributions: $M_p\sim(\gamma/d) h^2 \ell_I^{\, \infty,\text{open}}$, due to the depression inside the liquid column and $M_\gamma\sim \gamma h\ell_I^{\, \infty,\text{open}} $, induced by the pulling contact line at the top of the walls (see Methods section `Torque balance' for a detailed derivation).
For flexible enough sheets, contact between two neighbouring walls (case (iii)) occurs when the sheet radius of curvature $R=h$, yielding a bending moment $M_B \sim B/h$, where $B=Ee^3L/[12(1-\nu^2)]$ is the sheet bending stiffness. By choosing the proper width of the sheet, or equivalently the right number of grooves $n\approx 2\pi h /d$, this situation of self-contact between neighbouring walls corresponds to the closure of the whole structure into a tube. Thus, the formation of a tubular structure occurs when $M_B<M_p+M_\gamma$. In the limit of deep narrow grooves (i.e. $h\gg d$), as $M_p\gg M_\gamma$, this inequality becomes $Bd^2/\left[\gamma h^3\ell_c^2\right]<1$.
In the situation where $h\sim d$, the two contributions of the capillary torque are comparable and a more involved analysis is needed to derive the criterion for complete closure of the grooved sheet (see Methods section `Stationary solution and condition for groove closure'):
\begin{equation}
\label{lambda-def}
    \Lambda \equiv \frac{8Bd^2}{27\gamma h^3\ell_c^2}\left[1+2d/h+0.556\,(d/h)^2\right]^{-1} <1.
\end{equation}

When this criterion is fulfilled, the deformation of the groove reduces the channel cross-section and the equilibrium height of the first capillary rise increases to (see Methods section `Stationary solution and condition for groove closure'):
\begin{equation}
\label{ell-closed}
    \ell_I^{\, \infty,\text{closed}} = \frac{2\ell_c^2}{d}(2+d/h)=\ell^{\, \infty,\text{open}}_I\, (2+d/h),
\end{equation}
which is more than twice the elevation in the rectangular open groove (Eq.~(\ref{eq:lopen})).

Moreover, as the grooved sheet curls into a tube of inner radius $R_\text{int}=hw/d$~\cite{cappello2023bioinspired} determined by self contact between the neighbouring walls, the second capillary rise $\ell_{II}$ (Fig.~\ref{fig1}) reaches the classical Jurin height, $\ell_{II}^{\infty} = 2\ell_c^2/R_\text{int}$, at equilibrium (Methods section `Second capillary rise dynamics').
We can now compute the amount of liquid captured by the grooved sheet when it is rigid or flexible. For the rigid and flat case, the captured volume simply reads $V_R=nhd\ell^{\, \infty,\text{open}}_I$, where $n\approx 2 \pi h/d$ is the number of grooves. When the structure is flexible however, it closes and the grooves become triangular, leading to a volume captured in the grooves $V_F=nhd\ell^{\, \infty,\text{closed}}_I/2$. Using Eq.~(\ref{ell-closed}), we have $V_F=V_R (1+\bar{d}/2)$. Moreover, as the whole sheet curls and closes, an additional volume is trapped in the newly formed tube of radius $R_\text{int}=hw/d$, leading to an additional volume $V_C=\pi R_\text{int}^2\ell_{II}^{\, \infty}= V_R w/2h$ and hence a total captured volume increase of a factor $1+(w+d)/2h$ with respect to the rigid flat case. 

The validity of the criterion given by Eq.~(\ref{lambda-def}) can be assessed in Fig.~\ref{fig2}d where experimental measurements (symbols) of the equilibrium heights of the first (light blue triangles) and second (dark blue circles) capillary rise are shown as a function of the dimensionless transition parameter $\Lambda$. Indeed, experimental measurements show a sharp transition at $\Lambda\simeq1$ for both capillary rises $\ell_I$ (from $\ell^{\, \infty,\text{open}}_I$ to $\ell^{\, \infty,\text{closed}}_I$) and $\ell_{II}$ (from $0$ to $\ell_{II}^{\infty}$) between open and closed states.
Additionally, as shown in Fig.~\ref{fig2}e, the closure time $t_{\text{closure}}$ appears to diverge at the theoretically predicted transition. Although the precise geometry of the groove of a hummingbird tongue is quite different, the dimensionless number $\Lambda$ may be estimated thanks to data from the literature\cite{rico2011hummingbird,kim2012hummingbird} ($h\approx d\approx 150$ µm, $e\approx 25$ µm, $E\approx 300$ kPa, $L\approx 1$ cm, $\gamma\approx 0.06$ N/m, $\rho\approx 1000$ kg/m$^{3}$) and yields $\Lambda\approx 10^{-2}$, that is well inside the closing regime.

Having rationalised under which condition the first capillary rise may induce a closure of the structure, and hence, a second capillary rise, we now aim to describe the dynamics of this intricate fluid-structure interaction problem. Following the seminal reasoning of Washburn, we balance the driving capillary force with the resisting gravitational and viscous forces inside each groove of varying geometry due to elastic deformation. For the sake of simplicity, and based on our experimental observation, we assume that the sheet bends along an arc of circle and that elastic deformation reduces to a single degree of freedom, the closing angle $\beta$ of the groove (see Fig.~\ref{fig2}a), i.e., we consider that the deformation of the sheet is identical along the structure length. This assumption is valid as long as the curvature persistence length  -- defined as the length from which a curvature imposed at an extremity completely disappears -- is large compared to the length of the structure itself\cite{barois2014curved}. We also assume that the vertical air-liquid interface joining the wall free ends is bent along an arc of circle of radius $R-h$ with the same central angle 2$\beta$ as the sheet. The contact angle $\theta$ between the liquid and the walls is thus constant and equal to $\pi/2$ at any vertical elevation.\\
Simple energy considerations yields the driving capillary force $F_\gamma=2\gamma (1+\beta)h$ (see Methods section `Pressure difference in a groove'). The gravity force may be expressed as $F_g=-\rho g A(\beta,\bar{d})\ell_I$, where $\ell_I(t)$ is the elevation of the capillary rise in the groove, $\bar{d} = d/h$ is the groove aspect ratio and $A(\beta,\bar{d}) = h^2(\bar{d}-\beta)$ is the cross-section area of the groove. The derivation of the viscous friction induced by a Poiseuille-like flow in an open groove requires to solve the Stokes equations with proper boundary conditions, i.e., no-slip at liquid-solid interfaces and no-shear at liquid-air interfaces\cite{yang2011dynamics,kolliopoulos2021capillary}. The derivation is detailed in the Methods section `Flow in a groove' and yields a viscous friction force of the form $F_\mu=-\mu A^2(\beta,\bar{d}) G(\beta,\bar{d}){d}^{-4}\ell_Id\ell_I/dt$, where $\mu$ is the liquid viscosity and $G$ is a dimensionless function of $\beta$ and $\bar{d}$ related to the geometry of the groove. Balancing the three forces yields a non-linear differential equation for $\ell_I$. Yet, as there is two unknowns in the system ($\ell_I$ and $\beta$) we need a second equation to close the problem.\\ This additional equation is given by the mechanical moment-balance in the sheet. As discussed above, the elastic restoring moment $M_B = B\beta/(2d)$ is balanced by the two capillary contributions on the moment, induced by the depression $M_p=-h^2/2\int_0^{\ell_I}p(z)dz$, where the pressure is $p(z)=-2\gamma(1+\beta)/[d-h\beta]z/\ell_I$, and by the pulling triple line $M_\gamma=\gamma h\ell_I$ (Methods section `Torque balance'). To summarise, we obtain the following system of equation for $\beta$ and $\ell_I$:
\begin{align}
&2\gamma (1+\beta)h-\rho g A(\beta,\bar{d})\ell_I-\mu A^2(\beta,\bar{d}) {G}(\beta,\bar{d}){d}^{-4}\ell_I\frac{d\ell_I}{dt}=0\\
&B\frac{\beta}{2d}- \frac{1+\beta}{2[\bar{d}-\beta]}\gamma h\ell_I-\gamma h{\ell_I} =0,
\end{align}
together with the initial condition $\ell(0)=0$. Solving numerically this nonlinear system of equation yields the light blue line in Figs.~1d and 2c (see Methods section `Fluid-structure interaction and dimensionless equations' for more details), showing excellent agreement with the experiments within the intrinsic experimental uncertainties.  
For grooved sheets composed of $n\approx 2 \pi /\bar{d}$ grooves, contact between neighbouring walls corresponds to the deformation of the whole structure into a tube in which the second capillary rise, $\ell_{II}(t)$, takes place. Because self-contact induces a large increase in bending stiffness of the sheet, this second capillary rise follows the same dynamics as the one observed in a rigid tube\cite{zhmud2000dynamics,Fries2008}. Figures~\ref{fig1}d and \ref{fig2}c show excellent agreement between the model and the experiments for $\ell_{II}(t)$ (see also Methods section `Second capillary rise dynamics' and Supplementary Video 2). Varying the thickness $e$ and the length $L$ of the sheet, the equilibrium height of both capillary rises (Fig.~\ref{fig2}d) as well as the time for closure (Fig.~\ref{fig2}e) can be theoretically computed as a function of $\Lambda$ (see Methods sections `Stationary solution and condition for groove closure' and `Time needed for groove closure', respectively). 
Note that close to the closing transition, there is a large variability of the final capillary rise within the same sheet (Fig.~\ref{fig2}b, second picture and Fig.~\ref{fig2}d), which is not captured by our model, where each groove is assumed to behave in the same way. We interpret this large variability by the finite bending stiffness of the walls that experience a symmetry breaking event commonly reported in elastocapillary phenomena\cite{bradley2023bendocapillary,liu2021capillary,bico2018elastocapillarity}. Figure~\ref{fig2}e shows that the closure time of the structure $t_\text{closure}$ increases with $\Lambda$ and diverges at the transition (Supplementary Video 3). This increase is well captured by our model: a higher critical capillary rise is indeed needed to have enough force acting on the walls to bend the sheet when $\Lambda$ approaches 1, and reaching a higher elevation takes more time. 
The experimental critical value of $\Lambda$ at which $t_\text{closure}$ diverges is found to be slightly above 1 ($\approx 1.4$). This is due to a slight overestimation of the bending energy of the structure in the model which must be bent by an angle $\beta$ over its entire length, $L$, to close, whereas, in the experiments, it closes when it is bent over a portion of its entire length (see Fig.~\ref{fig1}c at $t=1$ s). At lower values of $\Lambda$, the slight discrepancy between theory and experiments is due to the small value of $t_\text{closure}$ for which inertial effects of the liquid become significant and are not taken into account in the model (See Methods section `Discussion on inertial effects').

\begin{figure}[!ht]
  \centering
    \includegraphics[width=0.7\textwidth]{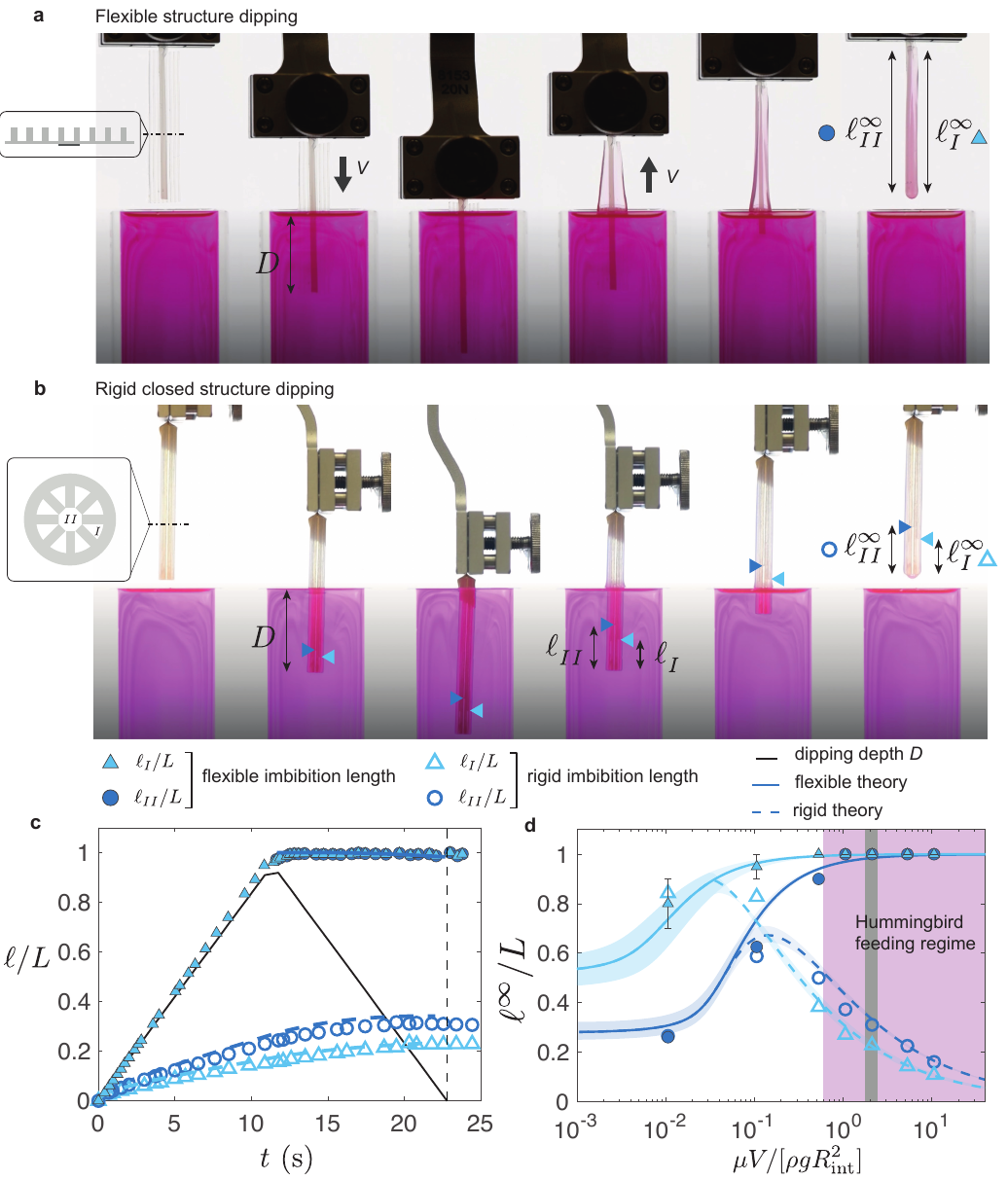}
       \caption{ {\bf Elasticity-enhanced fluid capture during dipping.} {\bf a}, Snapshots of a flexible grooved sheet as it is dipped and pulled out from a bath (silicon oil V1000, $\mu=0.96$ Pa.s, $\rho=960$ kg.m$^{-3}$, $\gamma=0.021$ N/m) at constant speed $V=200$ mm/min. Sheet geometry: $L=40$ mm, $h=800\pm20$ µm, $w=360\pm20$ µm, $d=700\pm50$ µm, $e=135\pm10$ µm. {\bf b}, Dipping of a rigid closed structure with the same geometry and parameters. The position of the imbibing liquid fronts both inside the peripheral triangles $\ell_I$ and the central tube $\ell_{II}$ are highlighted by blue arrows. {\bf c}, Rescaled imbibition length $\ell/L$ as a function of time during the dipping cycle shown in {\bf a} and {\bf b}; empty (rsp. filled) symbols correspond to the rigid (rsp. flexible) case. Dashed lines correspond to the impregnation theory (see Methods section `Full immersion of the structure into the liquid bath' for more details.  {\bf d}, Final imbibition length $\ell^{\, \infty}/L$ (indicated by the vertical dashed line in {\bf c}) as a function of the dimensionless number $\mu V/[\rho g R_\text{int}^2]$. Solid (rsp. dashed) lines correspond to the flexible (rsp. rigid) imbibition theory. Grey region indicates the data corresponding to {\bf c}. The hummingbird feeding regime is highlighted in purple.}
       \label{fig3}
\end{figure}

We now turn to the case where the liquid is abundant and the device is dynamically fully dipped into a bath. The key advantage of the flexibility of the device, but also of the  hummingbird tongue, resides in the amount of fluid captured in the process (see Supplementary Videos 4 and 5). Indeed, when the sheet is immersed in and then removed from a liquid bath at a speed $V$, the grooves are immediately filled with liquid during the protrusion phase (Fig.~\ref{fig3}a and c). During the retraction phase, surface tension induces the closure of the sheet into a tube, capturing the liquid which slowly drains from the closed grooves (Fig.~\ref{fig3}a). When a rigid structure with the same closed geometry (Fig.~\ref{fig3}b) is dipped into the bath, the penetration of the liquid inside the structure is much slower, as it has to flow through the cavity of the cylinder that has a huge hydraulic resistance because of its small size (Fig.~\ref{fig3}b and c). The driving force is the difference in hydrostatic pressure, that scales as $\rho g L$, whereas viscous shear, scaling as $\mu v L/R_\text{int}^2$ -- where $v$ is the typical penetrating front velocity -- resists the flow. Balancing both driving and resisting terms yields the typical penetrating speed $v\sim\rho g R_\text{int}^2/\mu$ that needs to be compared with the imposed dipping velocity $V$. For very slow dipping $V\ll v$, i.e., when $\mu V/[\rho g R_\text{int}^2]\ll 1$, the dipping is quasistatic, the fluid has the time to fully penetrate and then drain during the protrusion and retraction phases, respectively, leading to the same amount of liquid captured as in the static capillary rise (Fig.~\ref{fig2}d and~3d). Both rigid and flexible structures thus capture the same amount of fluid. For very fast dipping however, i.e. when $\mu V/[\rho g R_\text{int}^2]\gg 1$, the liquid does not have the time to penetrate inside the rigid closed structure, leading to a strong decrease in the captured volume, whereas, for the flexible ones, it penetrates the grooves with negligible friction during protrusion and barely drains during retraction, thus exhibiting a strong increase in the amount of liquid trapped (Fig.~\ref{fig3}c and d). The model, shown by the solid (rigid structures) and dashed lines (flexible structure) in Fig.~\ref{fig3}c and d, consists of a simple Lucas-Washburn law with a hydrostatic pressure at the groove entrance $p(z=0,t)$ varying with the dipping depth $D(t)$, $p(z=0,t) = \rho g D(t)$, where $D(t)=V t$ during immersion and $D(t) = L - V t$ during retraction (see Methods section `Full immersion of the structure into the liquid bath'). Applying biological data of the hummingbird tongue from the literature\cite{rico2011hummingbird,kim2012hummingbird} ($R\approx 0.1-0.2$ mm, $V\approx0.1-0.4$ m/s and $\mu\approx 10^{-3}-10^{-1}$ Pa.s) to our model, we deduce that hummingbird tongues are precisely in the regime where flexibility strongly promotes fluid capture compared to its rigid counterpart (purple region in Fig.~\ref{fig3}d).

\begin{figure}[!ht]
  \centering
    \includegraphics[width=0.9\textwidth]{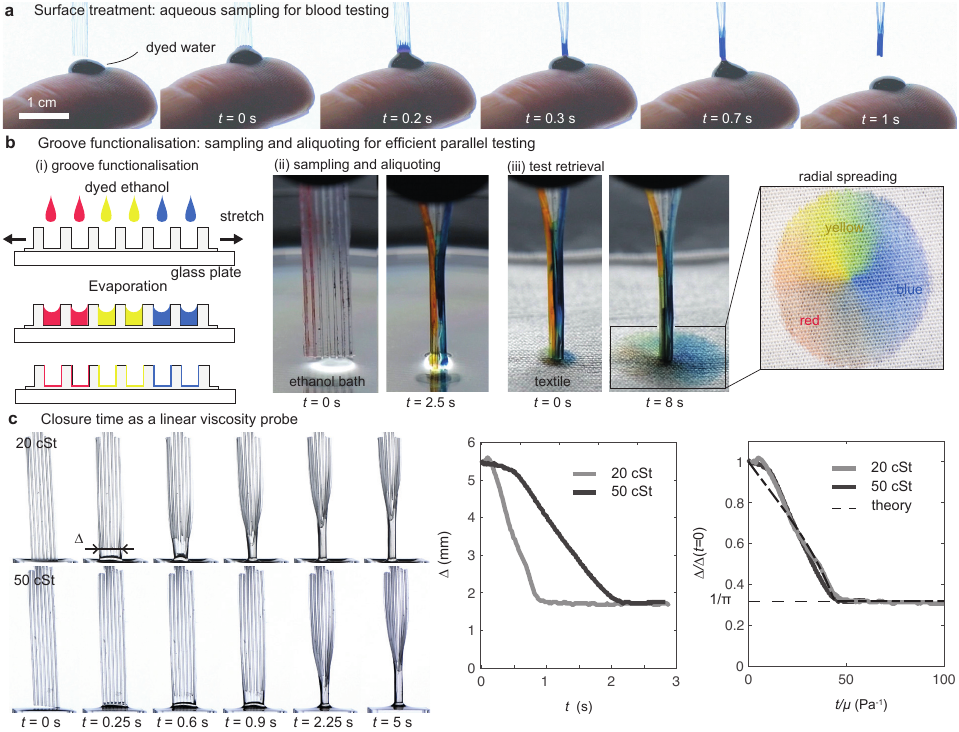}
       \caption{ {\bf All-in-one fluidic testing device.} {\bf a}, Thanks to its simple geometry, the device may be easily scaled down and plasma treated to swiftly capture a small fraction of a water drop. The initial open state enables a fast flow through the grooves and the closing of the structure impedes evaporation and unwanted exchanges with the environment ($h=w=d=300$ µm, $e=60$ µm). {\bf b}, Each groove may be additionally functionalised with specific markers -here, dye- using e.g. evaporation. Stretching the device and using the strong adhesion on glass, the grooves may be made transitorily larger to ease this step. When dipped in a fluid, the device swiftly samples and aliquots the liquid in each groove, where it efficiently reacts with the specific marker thanks to the triangular groove cross-section that ensures a large surface-to-volume ratio. The device is then put in contact with an absorbent textile, inducing the radial spreading of the captured liquid, enabling a direct reading of the role of the different markers.  {\bf c}, The closure time of the device acts as a linear probe for the viscosity of the liquid. The evolution of the apparent width $\Delta$ of the device may be collapsed by rescaling the time by the viscosity ($w=300\pm30$ µm, $d=500\pm30$ µm, $h=600\pm40$ µm, $e=110\pm10$ µm, $L=25.4$ µm). The dashed line correspond to the model  for  $h=600$ µm, $w=300$ µm, $d=500$ µm, $e=104$ µm, $L=25$ mm.}
       \label{fig4}
\end{figure}

In conclusion, this article demonstrates the efficacy for fluid capture of a flexible hierarchical structure composed of a set of vertical grooves that individually mimics the behaviour of a hummingbird's tongue. The intricate dynamics of concomitant fluid capture and structure deformation of the ``meta-tongue'' device, effectively described by a comprehensive analytical model that couples fluid flow at low Reynolds numbers with an elastica, occurs if  two essential conditions are fulfilled: first, the groove width must be smaller than the capillary length for capillary rise to occur, and second, the capillary forces must surpass the mechanical resistance of the lamellae beneath the grooves (i.e., $\Lambda<1$) to observe the structure deformation.
Hence, there is no limitation on the whole sheet width that can be much larger than the capillary length  while still closing into a tube.
From a mechanical perspective, similarly to the supporting rod in hummingbird tongues, the ribs contribute to the structural rigidity of the sheet in the dipping direction, thereby preventing viscosity-induced buckling of the structure as it enters the liquid bath. Furthermore, this configuration notably extends the curvature persistence length of the entire structure, a topic of interest for future research.
Comparative analysis of liquid capture by flexible structures versus rigid open or closed structures with identical texture geometry reveals that open rigid structures capture less liquid than flexible or closed rigid structures and that, when fully dipped into a liquid, fluid capture is notably more rapid with flexible structures than with closed rigid ones. 
Such comparison shows that evolution equipped hummingbirds with an ideal and versatile tool for liquid capture, that efficiently operates both when nectar is scarce (edge dipping and capillary rise) or abundant (full dipping and fluid trapping). Compared to the hummingbird tongue, our device goes one step further, as its 
hierarchical design, leading to the closure of the whole sheet into a tube, enables the capture of an additional volume of liquid.
Moreover, in contrast to rigid closed tubes, the flexible device possesses the remarkable ability to be initially flat, enabling straightforward storage, and may be reopened when compressed, simplifying the retrieval of the captured fluid.
Natural extensions of our work may harness the variation of the textures' geometries and the ability to design structures that curl into closed tubes with diverse and potentially more efficient cross-sectional geometries\cite{cappello2023bioinspired}, or on the design of structures with flexible walls that may collapse\cite{bradley2023bendocapillary} (see Supplementary Video 6).

Combining capillary suction and structure deformation thus opens new routes in passive capillary microfluidics\cite{Juncker2002,olanrewaju2018capillary}, especially for biological assays\cite{Narayanamurthy2020} and medical diagnostics\cite{tan2017glass} like blood testing\cite{Gao2020}, as it offers the opportunity to concomitantly collect and aliquot the sample in a fast and efficient manner.
This is demonstrated in Fig.~\ref{fig4}a,b in which sampling and aliquoting few microliters of an aqueous solution can be performed in a flexible structure, enabling the release in few seconds of pre-coated reagents (here, dyes) in the aliquots  (see Supplementary Video 7 and 8). In addition, contacting the tip of the imbibed structure to an absorbent substrate allows for a direct reading of the reaction (here, only diffusion -- see Supplementary Video 8). We believe that replacing the dye with dehydrated antibodies in the grooves should enable rapid blood testing using only a fraction of a drop volume instead of the three droplets needed for instance in a Serafol\textregistered \, ABO+D card used to confirm patient identity immediately before a blood transfusion. The reading could alternatively be done after reopening the structure and optically detecting the groove in which agglutination of the blood has occurred. Finally, Fig.~\ref{fig4}c and Supplementary Video 9 show that the closing time of the structure could rapidly inform on the hematocrit level, which is directly correlated to the blood viscosity\cite{Chien1966}. Our all-in-one fluidic testing device could  therefore compete with centrifugal microfluidics\cite{Strohmeier2015} for at least blood testing, as it is passive, faster, less expensive, and requires less volume of sample. These attributes are also compatible with those of paper-based microfluidics making medical diagnostics accessible for development countries\cite{Mao2012}.

\begin{addendum}
 \item We acknowledge support by F.R.S.-FNRS under the research grants CR n◦40017301 “G.El.In.Flow" and CR n◦40011004 “BioCapTure". This project has received funding from the European Union’s Horizon 2020 research and innovation program under the Marie Sklodowska-Curie grant agreement no. 101027862 and from 747 Fédération Wallonie-Bruxelles (ARC ESCAPE project). We thank the Micro-milli service platform (ULB) for the access to their experimental facilities.
 \item[Competing Interests] E.S., F. B., B. S. and J. C. are co-authors on a filed patent No. 24199680.0, which describes the methods used herein.
 \item[Correspondence] Correspondence and requests for materials should be addressed to E.S. and J.C.~(email: emmanuel.siefert@ulb.be and jean.cappello@ulb.be).
 \item[Authors contributions] E.S. and J.C. designed research; E.S. and J.C. performed research; E.S. and J.C. analyzed data; E.S., J.C. and F.B. developed the model; all authors wrote the manuscript.
 \item[Data and code availability] The raw data and the code supporting the findings of this study are available from the corresponding authors upon reasonable request.
\end{addendum}

\bibliography{apssamp}
\bibliographystyle{naturemag}


\setcounter{figure}{0}
\renewcommand{\figurename}{Extended Data Figure}

\methods

\subsection{Elastomer preparation.}

The structures are made of platinum-catalyzed silicone rubber (Ecoflex 00-50 from Smooth-On) and are fabricated by mixing a prepolymer base and a curing catalyst in a 1:1 weight ratio. Just after mixing, when the melt is still liquid, it is poured on a 3D-printed mold (printed with an Ultimaker S5) and spin-coated (SCS 6800 SpinCoater Series) at various rotation speeds that control the sheet thickness. The polymer melt gradually cures over time leading to an elastic solid. At room temperature curing time is 2 hours. After demoulding the sheets, they are placed in a bath of V10 silicon oil until they are completely swollen (typically overnight), leading to a uniform increase in lengths of $30\%$. This step enables us to use silicone oil as a completely wetting liquid (contact angle $\theta_Y=0$), without inducing any additional swelling of the sheet. The Young's modulus of the swollen elastomer is 57 kPa measured by mean of a traction machine (ZwickiLine  Z0.5TH  from Zwick) assuming material incompressibility (i.e., a Poisson ratio $\nu=0.5$).

\subsection{Experimental apparatus.}

\begin{figure}
  \centering
    \includegraphics[width=0.4\columnwidth]{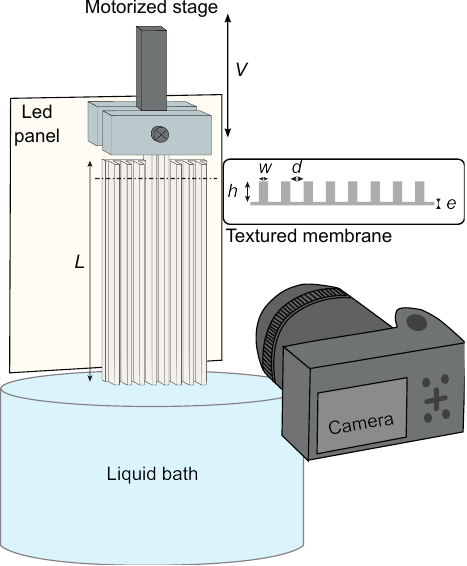}
      \caption{{\bf Schematics of the experimental setup.} A grooved sheet hanging from a motorised stage is dip in a wetting liquid bath. The dynamics of the capillary rise is monitored using a camera with a LCD backlight.
      }
      \label{fig_schem_exp_setup}
\end{figure}

A schematic of our experiments is shown in Extended Data Fig.~\ref{fig_schem_exp_setup}. 
The structured sheets are hanged by mean of a traction machine (ZwickiLine 0.5 kN from Zwick) above a liquid bath filled with silicone oil (V5, V10, V50 or V100 from Merck) with the grooves oriented  vertically. The structure is then either slightly dipped in the liquid reservoir or fully immersed in and withdrawn from the bath at a controlled velocity. While capillary rise occurs, the dynamics is recorded using a camera (Nikon D850) at a frame rate of 100 fps or 120 fps. The height of the first and second capillary rise are then extracted from the images by standard image processing using ImageJ and MatLab.
In the experiments involving the full immersion of the flexible sheet into the liquid bath, a lamella of Mylar of width comparable to the ribs width $d$, and of same length $L$ as the sheet, is fixed at the back of the structure to increase its structural rigidity in the dipping direction in order to prevent viscosity-induced buckling of the structures.

\subsection{Modelling approach.}

\begin{figure}[!t]
  \centering
    \includegraphics[width=\columnwidth]{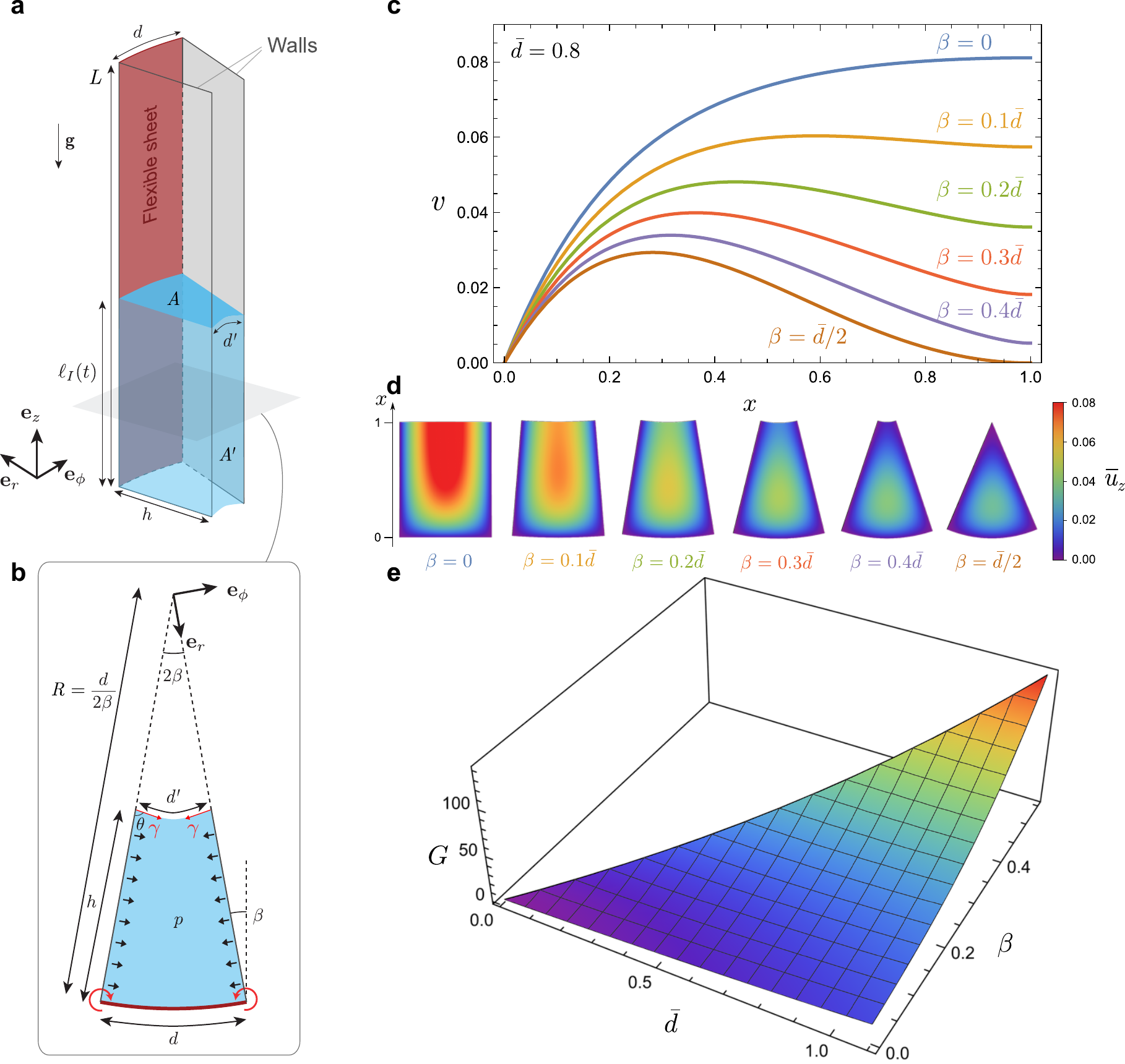}
     \caption{{\bf Modelling.} {\bf a}, Capillary rise with a contact angle $\theta=\pi/2$ in an open groove of depth $h$, width $d$ and making a closing angle $\beta$. {\bf b}, Schematic of the mechanical equilibrium: the capillary rise induces a torque through the walls (considered as rigid) due to the negative pressure and the surface tension pulling at the triple line, leading to the bending of the bottom sheet with a radius of curvature $R = d/(2\beta)$. {\bf c}, function $v$ (Eq.~(\ref{eq_barg})), representing the dimensionless mean vertical velocity $v_z$ as a function of $x=(R-r)/h$ for $\bar{d}=0.8$ and various values of $\beta=\sqrt{3}/k$. {\bf d}, Colourmaps of the dimensionless vertical velocity $\bar{u}_z =u_z\, \mu\ell_I/[(\Delta p-\rho g \ell_I) d^2]$ inside the grooves, given by Eqs.~(\ref{u-field-final}) and (\ref{vzbar}), for various values of $\beta$ and $\bar{d}=0.8$. {\bf e,} Plot of the dimensionless function $G$, defined by Eqs.~(\ref{Gtilde-def}) and (\ref{G-def}), as a function of the aspect ratio $\bar{d}=d/h$ and the closing angle $\beta$ of the groove.
     }
     \label{fig_schem}
\end{figure}

We start by presenting the general ideas of the model before giving the details in the next sections. To model the fluid-structure problem occurring when a fluid rises in a deformable grooved sheet, we consider the flow inside each groove, where the closing angle $\beta$ is the only degree of freedom in the deformation. The computation of the flow is based on the Washburn approach where it is assumed that the unidirectional flow is driven by the capillary pressure gradient, $\Delta p$, caused by the existence of a free surface, whereas viscous and gravity forces oppose to the flow\cite{washburn1921dynamics}. For the sake of simplicity, we neglect the inertial terms in the momentum equation, as they are significant only for the very early dynamics of the capillary rise and are thus irrelevant for our fluid-structure interaction study. Under these conditions, the flow rate $Q$ has a Poiseuille-like expression
\begin{equation}
\label{Qintro}
Q = \frac{1}{\mu}\left[\frac{\Delta p}{\ell_I}-\rho g\right]\, d^4\, G^{-1}(\beta,\bar{d}),
\end{equation}
where $\bar{d} = d/h$, $G$ is a function taking into account the geometry of the groove and $\ell_I$ is the liquid height in a single groove, see Extended Data Fig.~\ref{fig_schem}a. 

On the other hand, the flow rate is also equal to $Q=A(\beta,\bar{d}) d\ell_I/dt$, where $A$ is the area of an horizontal cross-section of the groove. Equating these two expressions gives an evolution equation for $\ell_I(t)$ once $\Delta p$ and $G$ are known. This equation involves $\beta$ which varies as the fluid rises. A second equation is thus needed to close the system. It is obtained from the balance of torques due to the bending of the sheet, the capillary pressure and the capillary force at the triple line. The fluid-structure problem is then solved by computing numerically the solution of these two coupled equations describing the simultaneous capillary rise inside the grooves, i.e. $\ell_I(t)$, and the bending of the sheet, i.e. $\beta(t)$. 

\subsection{Geometry of a groove and of the air-liquid interfaces.}
\label{sec:geometry}

A groove is composed of two rigid walls of height $h$ attached to a flexible sheet of width $d$ and having both a length $L$, see Extended Data Fig~\ref{fig_schem}a. When the fluid rises inside the groove, the walls move toward each others and their clamped ends rotate by an angle $\beta$ with respect to their initial position so that the distance between their free ends decrease. During this process, the flexible sheet bends. For simplicity, we assume that it bends along an arc of circle and that the walls stay perpendicular to the sheet during their rotation. The sheet and the walls form then a circular sector with a central angle $2\beta$ so that the radius of curvature of the sheet is given by (Extended Data Fig.~\ref{fig_schem}b)
\begin{equation}
\label{R-sheet}
R=\frac{d}{2\beta}.
\end{equation}
We also assume for simplicity that the vertical air-liquid interface $A'$ joining the wall free ends is bent along an arc of circle of radius $R-h$ with the same central angle $2\beta$. The contact angle $\theta$ between the liquid and the walls is thus constant and equal to $\pi/2$ in this case. At any vertical elevation $z$, the length $d'$ of this interface is thus
\begin{equation}
\label{dp-eq}
d' = 2 \beta (R-h) = d- 2h\beta, \quad 0\le \beta \le d/(2h).
\end{equation}
When $\beta = d/(2h)$, free ends of the walls are in contact and $d'=0$. The area of the vertical air-liquid interface is thus given by
\begin{equation}
\label{Ap-eq}
A'(\beta,\bar{d}) = d'\, \ell_I.
\end{equation}

Neglecting the curvature of the horizontal air-liquid interface $A$, its area coincide with the area $A$ of an horizontal cross-section of a groove given by the difference between the area of two circular sectors of radius $R$ and $R-h$ with the same central angle $2\beta$:
\begin{equation}
\label{A-eq}
A(\beta,\bar{d}) = \beta R^2 - \beta (R-h)^2 = h^2\, (\bar{d}-\beta) \equiv h^2\, \bar{A}(\beta,\bar{d}),
\end{equation}
where $\bar{d} = d/h$.

In this simplified geometry, the centres of curvature of both the sheet and the vertical air-liquid interface $A'$ are located on the same vertical line where the $z$-axis of a cylindrical system of coordinates will be placed to compute the flow inside the groove in the next section. Doing so, the boundaries of a groove are described by coordinate lines allowing to separate variables when solving the Stokes equations. 


\subsection{Flow in a groove.}

The flow inside a groove is described in cylindrical coordinates $({\bf e}_r,{\bf e}_\phi,{\bf e}_z)$ whose origin is located at the bottom of the groove on the centre of curvature of the flexible sheet, see Extended Data Fig.~\ref{fig_schem}b. The groove is thus described by $(r,\phi,z$) with $R-h \le r \le R$, $-\beta \le \phi \le \beta$ and $0\le z \le \ell_I$ and its cross-sectional area is given by Eq.~(\ref{A-eq}). We assume the flow to be unidirectional along the $z$-direction, so that the $r$ and $\phi$ component of the velocity field ${\bf u}$ are vanishing ($u_r=u_\phi=0$), and the fluid to be an incompressible Newtonian fluid of density $\rho$ and viscosity $\mu$. In this case, mass conservation imposes the following form for ${\bf u}$:
\begin{equation}
\label{u-field1}
    \frac{\partial u_z}{\partial z}=0 \Rightarrow {\bf u}=u_z(r,\phi){\bf e}_z.
\end{equation}
The Navier-Stokes equations in cylindrical coordinates simplifies to:
\begin{equation}
\rho \frac{\partial u_z}{\partial t}= -\rho g -\frac{\partial p}{\partial z}+\mu \nabla^2 u_z, \quad \frac{\partial p}{\partial r}= \frac{\partial p}{\partial \phi} =0,
\label{eq-NS}
\end{equation}
where $g$ is the gravitation acceleration, i.e. ${\bf g} = -g\, {\bf e}_z$. Because $u_z$ does not depend on $z$, taking the $z$-derivative of Eq.~(\ref{eq-NS}) leads to $\partial_z(\partial_z p) = 0$ so that:
\begin{equation}
\label{pressure-eq}
p(z)=p_{\text{atm}}-\frac{\Delta p}{\ell_I}\,z,
\end{equation}
where $p_{\text{atm}}$ is the atmospheric pressure, $\Delta p$ is the pressure difference between the entry ($z=0$) and the exit of the groove ($z=\ell_I$) and $\ell_I$ is the length of the groove filled by the liquid. Searching for stationary solution and writing the Laplacian operator in cylindrical coordinates, the first of Eqs.~(\ref{eq-NS}) simplifies to:
\begin{equation}
    \frac{\partial^2u_z}{\partial r^2}+\frac{1}{r}\frac{\partial u_z}{\partial r}+\frac{1}{r^2}\frac{\partial^2u_z}{\partial \phi^2}+\frac{1}{\mu}\left[\frac{\Delta p}{\ell_I}-\rho g\right]=0.
    \label{eq:NS-s}
\end{equation}
Due to the non-slip boundary condition, the velocity vanishes along walls:
\begin{equation}
    u_z(r,\phi=\pm \beta)=0.
\end{equation}
It is usual to use the following quadratic ansatz to satisfy those two boundary conditions\cite{nasto2018viscous,wei2022trade}:
\begin{equation} 
u_z(r,\phi)=v_z(r)\, \varphi(\phi)=v_z(r)\, \frac{3(\beta^2-\phi^2)}{2\beta^2}.
    \label{eq:ansatz}
\end{equation}
Substituting Eq.~(\ref{eq:ansatz}) into Eq.~(\ref{eq:NS-s}), we obtain
\begin{equation}
    \varphi\, \frac{d^2v_z}{d r^2}+\frac{\varphi}{r}\frac{d v_z}{d r}+\frac{v_z}{r^2}\frac{d^2 \varphi}{d\phi^2}+\frac{1}{\mu}\left[\frac{\Delta p}{\ell_I}-\rho g\right]=0.
    \label{eq:NS-sep}
\end{equation}
We now compute the mean flow along $z$ by taking the average along the $\phi$ direction. Noting that
\begin{equation}
\label{integ-f}
    \frac{1}{2\beta}\int_{-\beta}^{\beta}\varphi(\phi)\, d\phi=1,\quad \frac{1}{2\beta}\int_{-\beta}^{\beta}\frac{d^2 \varphi}{d\phi^2}\, d\phi=-\frac{3}{\beta^2},
\end{equation}
the $\phi$-average of Eq.~(\ref{eq:NS-sep}) reads:
\begin{equation}
     \frac{d^2v_z}{d r^2}+\frac{1}{r}\frac{d v_z}{d r}-\frac{3}{\beta^2}\frac{v_z}{r^2}+\frac{1}{\mu}\left[\frac{\Delta p}{\ell_I}-\rho g\right]=0.
     \label{eq:NSODE}
\end{equation}
Thanks to this ansatz, variables have been separated and we are left with an ordinary second-order differential equation (\ref{eq:NSODE}) for $v_z$ which requires two boundary conditions to be solved. The no-slip condition at the flexible sheet gives a first boundary condition:
\begin{equation}
\label{BC-1}
    v_z(r=R)=0,
\end{equation}
where $R$ given by Eq.~(\ref{R-sheet}) is the radius of curvature of the sheet. At the air-liquid interface, that we assume to be at a constant $r= R-h$ as we did to compute $A$ [Eq.~(\ref{A-eq})], we impose a no-stress boundary condition:
\begin{equation}
\label{BC-2}
    v'_z(r=R_c=R-h)=0.
\end{equation}


The solution of Eq.~(\ref{eq:NSODE}) together with the boundary conditions (\ref{BC-1}) and (\ref{BC-2}) reads:
\begin{subequations}
\label{vzbar}
\begin{align}
\label{vz}
    v_z(\bar{r})&=\frac{d^2}{\mu}\left[\frac{\Delta p}{\ell_I}-\rho g\right]v(\bar{r},k,\bar{R}_{c}),\\
    v&=c_1\bar{r}^k+c_2\bar{r}^{-k}+\frac{k^2\bar{r}^2}{12\, (k^2-4)},\label{eq_barg}\\
    c_1&=-\frac{k\, (k+2\bar{R}_{c}^{k+2})}{12\, (k^2-4)\, (1+\bar{R}_{c}^{2k})},\\
    c_2&=\frac{k\, \bar{R}_{c}^{k}\, (2 \bar{R}_{c}^{2}-k \bar{R}_{c}^{k})}{12\, (k^2-4)\, (1+\bar{R}_{c}^{2k})},
\end{align}
\end{subequations}
where $\bar{r}=r/R$, $\bar{R_c}=R_c/R= 1-h/R$ and $k=\sqrt{3}/\beta$. This solution is valid provided $k\not= 2$. An exact solution exists also for $k=2$ but we do not write it here because we assume $d/h \lesssim 1$ so that $k>2$ because $\beta \le d/(2h)$. Note that, using Eq.~(\ref{R-sheet}), we also have $\bar{R_c}=1-2\beta/\bar{d}$ with $\bar{d}=d/h$. In addition, $r$ varies between $R-h$ and $R$ so that $x = (R-r)/h=R/h (1-\bar{r})$ varies between $0$ when $r=R$ and $1$ when $r=R-h$.  Therefore, the dimensionless mean vertical velocity $v$ can be plotted as a function of $x$ with $\bar{d}$ and $\beta$ as parameters (instead of $k$ ans $\bar{R}_{c}$) as shown in Extended Data Fig.~\ref{fig_schem}c for $\bar{d}=0.8$. Indeed, for a given groove, $\bar{d}$ is a constant and only $\beta$ varies during the fluid rise whereas both $k$ and $\bar{R}_{c}$ varies with $\beta$. 

Finally, the complete velocity field is obtained from Eqs.~(\ref{u-field1}) and (\ref{eq:ansatz}):
\begin{equation}
\label{u-field-final}
    {\bf u}= u_z(r,\phi)\, {\bf e}_z = 3\, v_z(r)\left[\frac{\beta^2-\phi^2}{2\beta^2}\right]\, {\bf e}_z,
\end{equation}
with $v_z$ given by Eq.~(\ref{vzbar}). This velocity field is shown in Extended Data Fig.~\ref{fig_schem}d for $\bar{d}=0.8$ and several values of $\beta$.

To obtain an evolution equation for $\ell_I$ as a function of $\beta$ and $\bar{d}$, we now compute the flow rate $Q$
\begin{equation}
Q=\int_{-\beta}^{\beta}\int_{R_c}^{R}r \, u_z(r,\phi)\, d\phi\, dr=2\beta R^{2}\int_{\bar{R}_{c}}^{1}\bar{r}\, v_z(\bar{r}) d\bar{r}.
\end{equation}
where we used the change of variable $\bar{r}=r/R$ and Eq.~(\ref{integ-f}) to compute the integral over $\phi$. Using Eqs.~(\ref{R-sheet}) and (\ref{vz}), we have
\begin{subequations}
\begin{align}
\label{Q1}
Q &=\frac{d^4}{\mu}\left[\frac{\Delta p}{\ell_I}-\rho g\right]\, \tilde{G}^{-1}(k,\bar{R}_{c}), \\
\tilde{G}^{-1} &= \frac{k}{2\sqrt{3}}\int_{\bar{R}_{c}}^{1}\bar{r}\, v(\bar{r},k,\bar{R}_{c}) d\bar{r},
\end{align}
\end{subequations}
where $\beta=\sqrt{3}/k$ have been used. The remaining integral can be computed leading to
\begin{align}
\label{Gtilde-def}
&\tilde{G}(k,\bar{R}_{c}) = \left[96\sqrt{3}\, k^{-2}\,(k^2-4)^2\, (1+\bar{R}_{c}^{2k})\right]\times \nonumber \\
& \left[k(k-2)^2-(k-4)(k+2)^2\, \bar{R}_{c}^4 + k(k+2)^2 \, \bar{R}_{c}^{2k} \right. \nonumber \\
& \left.-32 k\, \bar{R}_{c}^{2+k}-(k+4)(k-2)^2 \, \bar{R}_{c}^{4+2k}\right]^{-1}.
\end{align}
Again, the parameters of interest in our system are $\bar{d}$ and $\beta$ instead of $k$ and $\bar{R}_{c}$. We thus define
\begin{subequations}
\label{G-def}
\begin{align}
G(\beta,\bar{d}) &= \tilde{G}(k(\beta), \bar{R}_{c}(\beta,\bar{d})), \\ 
k&=\frac{\sqrt{3}}{\beta}, \quad \bar{R}_{c} = 1-\frac{2\beta}{\bar{d}}.
\end{align}
\end{subequations}
Despite its apparent complexity, the behaviour of the function $G$ is quite simple in the regime of interest as shown in Extended Data Fig.~\ref{fig_schem}e, i.e. for $0\leq\bar{d}\leq 1.1$ and for $0\leq\beta\leq\bar{d}/2$ (upper bound corresponding to contact between neighbouring walls).


The evolution equation for $\ell_I$ is obtained by equating Eq.~(\ref{Q1}) to its equivalent expression, namely $Q=A(\beta,\bar{d})d\ell_I/dt$ where $A$ is given by Eq.~(\ref{A-eq}):
\begin{equation}
\label{eq-ell-temp}
\frac{\mu}{d^4} A(\beta,\bar{d})\, G(\beta,\bar{d})\, \ell_I\, \dot{\ell_I} = \Delta p-\rho g \ell_I,
\end{equation}
where $\dot{\ell_I} = d\ell_I/dt$. We thus need now to compute $\Delta p$ which is the remaining unknown quantity in this equation.

\subsection{Pressure difference in a groove.}

The pressure difference is due to the driving capillary force which can be derived from the change $dF$ in the free energy of the system when the liquid moves from an height $\ell_I$ to an height $\ell_I + d\ell_I$ in the groove. The free energy reads as
\begin{subequations}
\begin{align}
F &= \gamma\, A_{LV}+\gamma_{SL}\, A_{SL}+\gamma_{SV}\, A_{SV}, \\
A_{LV} &= A(\beta,\bar{d}) + A'(\beta,\bar{d}), \quad A_{SL} = (2h + d)\, \ell_I, \\
A_{SV} &= (2h + d)\, (L-\ell_I), \quad \gamma_{SV} - \gamma_{SL} = \gamma \cos \theta_Y,
\end{align}
\end{subequations}
where $A_{LV}$, $A_{SL}$, and $A_{SV}$ are the areas of the liquid-vapor, solid-liquid, solid-vapor interfaces, respectively, $\gamma \equiv\gamma_{LV}$, $\gamma_{SL}$, and $\gamma_{SV}$ are the energy per unit area of the liquid-vapor, solid-liquid, solid-vapor interfaces, respectively, $A(\beta,\bar{d})$ is given by Eq.~(\ref{A-eq}), $A'(\beta,\bar{d})$ is given by Eq.~(\ref{Ap-eq}) and the last relation is the Young's equation. Considering $\theta_Y=0$, the expression of $F$ becomes
\begin{equation}
F = \gamma \left[h^2 (\bar{d}-\beta) - 2(1+\beta) h \, \ell_I \right] + \gamma_{SV}(2h+d)L.
\end{equation}
The capillary pressure in then given by
\begin{equation}
\label{deltap-eq}
\Delta p = -\frac{1}{A} \frac{dF}{d\ell_I} =\frac{2\gamma (1+\beta)h}{A(\beta,\bar{d})} = \frac{2\gamma (1+\beta)}{h(\bar{d}-\beta)}.
\end{equation}

Substituting Eq.~(\ref{deltap-eq}) in Eq.~(\ref{eq-ell-temp}) and multiplying both side by $A$, we obtain the evolution equation for liquid height $\ell_I$ in the groove
\begin{equation}
\label{eq-ell-dim}
\frac{\mu}{d^4} A^2(\beta,\bar{d})\, G(\beta,\bar{d})\, \ell_I\, \dot{\ell_I} = 2\gamma (1+\beta)h -\rho g A(\beta,\bar{d}) \ell_I.
\end{equation}
Note that this equation corresponds to the force balance mentioned in the article  : 
\begin{equation}
F_\mu+F_\gamma +F_g =0,
\end{equation}
where $F_\mu = -{\mu} A^2(\beta,\bar{d})\, G(\beta,\bar{d})\, \ell_I\, \dot{\ell_I}/{d^4}$ is the viscous resistive force, $F_\gamma = 2\gamma (1+\beta)h$ is the driving capillary force and $F_g =  -\rho g A(\beta,\bar{d}) \ell_I$ is the gravitational force.
Equation~(\ref{eq-ell-dim}) depends on $\beta$ which varies during the liquid rise, hence, we need a second equation relating $\ell_I$ and $\beta$, to get a closed system of equations. It is obtained from the torque balance.

\subsection{Torque balance.}

The flexible sheet in between neighbouring walls is subjected to a bending moment due to both the depression in the rising liquid column $M_p$ and the pulling contact line at the top of the walls $M_\gamma$. Hence, neglecting the sheet inertia, the elastic restoring torque $M_B$ is given at every instant $t$ by:
\begin{equation}
M_B=\frac{B}{R}=M_p+M_\gamma,
\label{torque_balance}
\end{equation}
where $B=E e^3 L/[12(1-\nu^2)]$ is the bending modulus of the flexible sheet, $E$ and $\nu$ the Young modulus and the Poisson ratio of the material, respectively, and $R=d/(2\beta)$ the radius of curvature of the sheet. The moment due to depression in the groove reads
\begin{equation}
\label{Mp}
M_p=\frac{h^2}{2}\int_0^{\ell_I}(p_{\text{atm}}-p(z))dz = \frac{h^2 \Delta p}{2 \ell_I} \int_0^{\ell_I}z \, dz,
\end{equation}
where $p(z)$ is given by Eq.~(\ref{pressure-eq}) and $\Delta p$ by Eq.~(\ref{deltap-eq}). We thus obtain
\begin{equation}
M_p = \frac{1+\beta}{2(\bar{d}-\beta)}\gamma h\ell_I.
\end{equation}

The moment induced by the surface tension at the contact line is given by
\begin{equation}
\label{Mgamma}
M_\gamma=\gamma h\int_0^{\ell_I} \sin{\theta}\, dz = \gamma h \ell_I.
\end{equation}
where $\theta=\pi/2$ is the angle between the vertical air-liquid interface and the walls. 

The second equation relating $\beta$ to $\ell_I$ is obtained by substituting Eqs.~(\ref{Mp}) and (\ref{Mgamma}) in Eq.~(\ref{torque_balance})
\begin{equation}
\label{eq-beta-dim}
\frac{2\beta B}{d} = \gamma h \ell_I \left[1+ \frac{1+\beta}{2(\bar{d}-\beta)}\right].
\end{equation}

\begin{figure*}[!t]
  \centering
    \includegraphics[width=\textwidth]{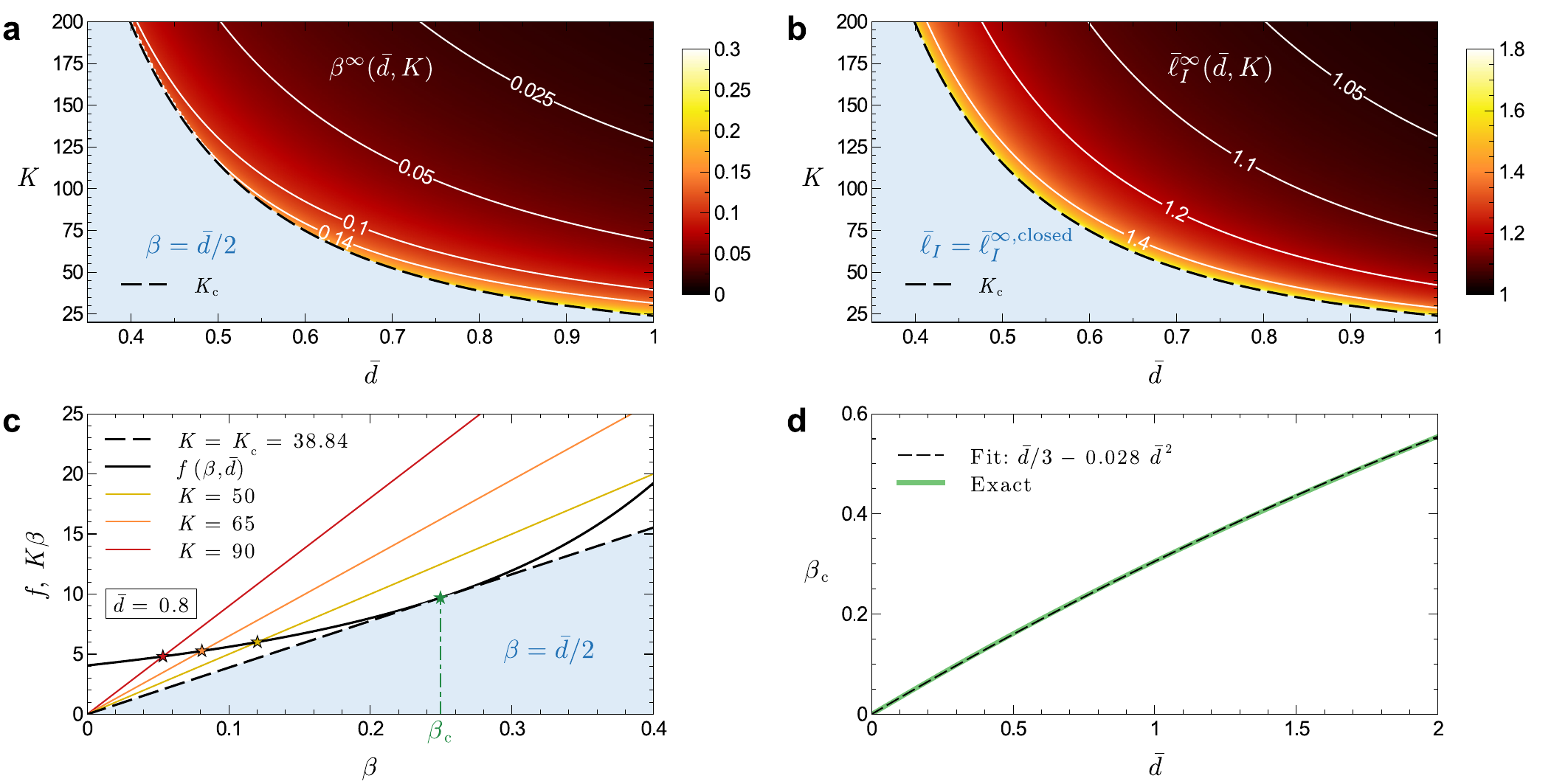}
      \caption{Evolution of the equilibrium value of the closing angle, $\beta^{\infty}$ defined by Eq.~(\ref{eq-betaJ}), ({\bf a}) and of the liquid height in a groove, $\bar{\ell}_I^{\, \infty}$, as a function of $\bar{d}$ and $K$. The critical value $K=K_c$, given by Eq.~(\ref{Kc}), below which the groove is closed is also shown. {\bf c.} Plots of the right-hand side of Eq.~(\ref{eq-betaJ}) (black solid line) together with the linear elastic relation on the left-hand side as a function of $\beta$ ($\bar{d}=0.8$), for various values of the dimensionless stiffness $K$. Intersection between the two curves, highlighted by stars, yields the static closing angle $\beta$ at equilibrium. Below a critical value $K_c$, given by Eq.~(\ref{Kc}), of the dimensionless stiffness, no solutions exist, and the groove closes until self-contact. {\bf d.} Evolution of $\beta_c$, given by Eq.~(\ref{betac}), as a function of $\bar{d}$ together with a polynomial approximation.}
       \label{fig_betaJ}
\end{figure*}

\subsection{Fluid-structure interaction and dimensionless equations.}

The fluid-structure problem where a liquid rises in a deformable groove is addressed by solving the coupled Eqs.~(\ref{eq-ell-dim}) and (\ref{eq-beta-dim}). For this purpose, it is convenient to use dimensionless equations. We rescale the variables as follows
\begin{equation}
\label{adim}
\bar{d} = d/h, \quad \bar{A} = A/h^2, \quad \bar{\ell_I} = \ell_I/\ell_I^{\, \infty,\text{open}}, \quad \bar{t} = t/\tau_I,
\end{equation}
where 
\begin{equation}
\ell_I^{\, \infty,\text{open}}=\frac{2\ell_c^2}{d}, \quad \ell_c = \sqrt{\frac{\gamma}{\rho g}},
\end{equation}
are the stationary height of the liquid in a groove with $\beta = 0$ (as shown in Sec.~`Stationary solution and condition for groove closure') and the capillary length, respectively, and where $\tau_{I}$ is given by
\begin{equation}
    \tau_{I}=\frac{\mu}{\rho g}\frac{\ell_c^2h^3}{d^6}.
\end{equation}
With these change of variables, Eqs.~(\ref{eq-ell-dim}) and (\ref{eq-beta-dim}) become
\begin{subequations}
\label{model-adim}
\begin{align}
\label{eq-ell-adim}
& 2 (\bar{d}-\beta)^2\, G(\beta,\bar{d})\, \bar{\ell_I}\, \dot{\bar{\ell_I}} = (1+\beta) -\left[1-\frac{\beta}{\bar{d}}\right] \bar{\ell_I}, \\
\label{eq-beta-adim}
& K \, \beta = \left[2+ \frac{1+\beta}{(\bar{d}-\beta)}\right] \frac{\bar{\ell_I}}{\bar{d}},
\end{align}
\end{subequations}
where $K=2B/[\gamma d\ell_c^2]$ is the dimensionless stiffness of the sheet and $G$ is defined by Eqs.~(\ref{Gtilde-def}) and (\ref{G-def}). This nonlinear system of equations can be solved numerically using Matlab or Mathematica with the initial condition $\bar{\ell_I} = 0$ at $\bar{t}=0$.

In the next sections, we discuss the stationary solution and the short-time dynamics as well as the condition on $K$ to reach a contact between neighbouring walls and the time it takes when it happens. 

\subsection{Stationary solution and condition for groove closure.}
\label{sec:stationary}

The stationary solution is obtain by setting $\dot{\bar{\ell_I}}=0$ in Eq.~(\ref{eq-ell-adim}). This leads to
\begin{equation}
\label{ell-J}
\bar{\ell}_I^{\, \infty} = \frac{(1+\beta)\bar{d}}{(\bar{d}-\beta)} \quad \Rightarrow \quad \ell_I^{\, \infty} = \frac{2\ell_c^2}{h}\frac{(1+\beta)}{\bar{d}-\beta}.
\end{equation}
Therefore, $\ell_I^{\, \infty}$ varies between $\ell_I^{\, \infty,\text{open}}$ and $\ell_I^{\, \infty,\text{close}}$ when $\beta$ varies between 0 and $\bar{d}/2$ where 
\begin{equation}
\ell_I^{\, \infty,\text{open}}=\frac{2\ell_c^2}{d}, \quad \ell_I^{\, \infty,\text{closed}}=\frac{2\ell_c^2}{d}(2+\bar{d}).
\end{equation}
The value $\beta^{\infty}$ of $\beta$ corresponding to $\bar{\ell_I}=\bar{\ell}_I^{\, \infty}$ (and which fully determines $\bar{\ell}_I^{\, \infty}$ once substituted in Eq.~(\ref{ell-J})) is obtained by replacing $\bar{\ell_I}$ by $\bar{\ell}_I^{\, \infty}$ in Eq.~(\ref{eq-beta-adim})
\begin{equation}
\label{eq-betaJ}
K \, \beta^{\infty} = \left[2+ \frac{1+\beta^{\infty}}{(\bar{d}-\beta^{\infty})}\right] \frac{(1+\beta^{\infty})}{(\bar{d}-\beta^{\infty})} \equiv f(\beta^{\infty},\bar{d}).
\end{equation}
This is a third order algebraic equation for $\beta^{\infty}$ which can be solved exactly. However, the expression of the solutions, which depends on $\bar{d}$ and $K$, is cumbersome and is not written here as we are mainly interested by the case where the neighbouring walls are in contact, i.e. when $\beta = \bar{d}/2$, as discussed below. 

Extended Data Figure~\ref{fig_betaJ}a shows a colour map and contour plot of $\beta^{\infty}$ as a function of $\bar{d}$ and $K$ and Extended Data Fig.~\ref{fig_betaJ}b shows similar plots for $\bar{\ell}_I^{\, \infty}$ obtained by substituting $\beta^{\infty}$ in Eq.~(\ref{ell-J}). For a given value of $\bar{d}$, say $0.8$, the value of $\beta^{\infty}$ decreases to $0$ and $\bar{\ell}_I^{\, \infty}$ tends to $1$ as $K$ increases. Indeed, $K$ measures the stiffness of the sheet relatively to capillary forces. When $K$ is large, the flexible sheet is stiff and barely bends ($\beta$ stays close to 0) and, consequently, $\bar{\ell}_I^{\, \infty}$ tends to the equilibrium height in an open groove, i.e. $\ell_I^{\, \infty} = \ell_I^{\, \infty,\text{open}}$ and thus $\bar{\ell}_I^{\, \infty}=1$, see Eq.~(\ref{adim}). 

As $K$ decreases, both $\beta^{\infty}$ and $\bar{\ell}_I^{\, \infty}$ increase until $K$ reaches a critical value $K_c$ below which there is no solution for Eq.~(\ref{eq-betaJ}) in the physical interval $0\le \beta^{\infty} \le \bar{d}/2$. Indeed, the function $f(\beta^{\infty},\bar{d})$ defined in Eq.~(\ref{eq-betaJ}) is singular at $\beta = \bar{d}$ and has thus two branches. When $K<K_c$, Eq.~(\ref{eq-betaJ}) has only one real solution corresponding to the intersection between $K\beta$ and the second branch of $f$ located at $\beta > \bar{d}$. Since $\beta$ cannot be larger than $\bar{d}/2$, we have $\beta=\bar{d}/2$ when $K<K_c$. We do not add a superscript $\infty$ to $\beta$ in this case since this extreme value of $\beta$ does not correspond to a mechanical equilibrium and is thus not a solution of Eq.~(\ref{eq-betaJ}).

The existence of a critical value of $K$ is illustrated in Extended Data Fig.~\ref{fig_betaJ}c where both functions $K\beta$ and $f(\beta,\bar{d})$ are plotted for $\bar{d}=0.8$ and several values of $K$ as a function of $\beta$ limited to its physical interval $[0,\bar{d}/2]$. The solution of Eq.~(\ref{eq-betaJ}), for a given $K$, is the intersection between these two functions of $\beta$ and are shown as star symbols. When $K<K_c$, there is no longer any intersection between these functions within the physical interval and $\beta$ jump from $\beta_c(\bar{d}) = \beta^{\infty}(\bar{d},K_c(\bar{d}))$ to $\beta = \bar{d}/2$ since $\beta^{\infty}(\bar{d},K<K_c)$ becomes larger than $\bar{d}/2$.  

The critical value of $K$ is thus such that the linear function $K_c\beta$ is tangent to the function $f(\beta,\bar{d})$, 
\begin{equation}
\label{eq-betac-Kc1}
\frac{\partial f}{\partial \beta}\Big|_{\beta=\beta_c} = K_c.
\end{equation}
Also, Eq.~(\ref{eq-betaJ}) gives the following relation for $K_c$:
\begin{equation}
\label{eq-betac-Kc2}
K_c \, \beta_c = f(\beta_c,\bar{d}).
\end{equation}
Eliminating $K_c$ from Eqs.~(\ref{eq-betac-Kc1}) and (\ref{eq-betac-Kc2}), we get an equation for $\beta_c$
\begin{equation}
\label{eq-betac}
\frac{\partial f}{\partial \beta}\Big|_{\beta=\beta_c} = \frac{f(\beta_c,\bar{d})}{\beta_c}
\end{equation}
where $f$ is given by Eq.~(\ref{eq-betaJ}). This is a third order algebraic equation for $\beta_c$ which possesses only one solution in the physical interval for $\beta$, i.e. $[0,\bar{d}/2]$. This solution depends only on $\bar{d}$, in contrast to the solution of Eq.~(\ref{eq-betaJ}) for $\beta^{\infty}$ which depends on both $\bar{d}$ and $K$, and it can be express in a rather compact form as follow
\begin{subequations}
\label{betac}
\begin{align}
\beta_c(\bar{d}) &= \bar{d}-(1+\bar{d})\left[\cos\sigma(\bar{d}) - \sqrt{3} \sin \sigma(\bar{d})\right]\approx \frac{\bar{d}}{3}-0.028\, \bar{d}^{\, 2}, \\ 
\label{sigma}
\sigma(\bar{d}) &= \frac{1}{3}\arctan\left(\sqrt{1+2\bar{d}}/\bar{d}\right).
\end{align}
\end{subequations}
This function is shown in Extended Data Fig.~\ref{fig_betaJ}d and is well approximated by a simple polynomial for $\bar{d}$ not too large. The function $K_c$ is then obtained by substituting $\beta_c$ in Eq.~(\ref{eq-betac-Kc1}) or (\ref{eq-betac-Kc2}):
\begin{equation}
\label{Kc}
K_c(\bar{d}) = \frac{2}{(1+\bar{d})}\left[\cos\sigma(\bar{d}) - \sqrt{3} \sin \sigma(\bar{d})\right]^{-3} \approx \frac{27}{4 \bar{d}^3\left(1+ 2\bar{d}+0.556\, \bar{d}^{\, 2}\right)},
\end{equation}
where $\sigma$ is defined in Eq.~(\ref{sigma}). The function $K_c$ is plotted in both Extended Data Fig.~\ref{fig_betaJ}a and b. This expression of $K_c$ provides a simple necessary condition for the groove closure, i.e. $K < K_c$. Using the definition of $K$, this criteria can be written as
\begin{equation}
\label{Bc}
B < B_c = \frac{\gamma h \ell_c^2}{2}\, \bar{d}\, K_c(\bar{d}).
\end{equation}
For a given liquid ($\gamma$) and a given depth ($h$) and width ($d$) of the groove, Eq.~(\ref{Bc}) gives the value of bending modulus of the flexible sheet below which neighbouring walls are brought into contact (the groove closes) when the liquid rises.
Using the approximate expression of $K_c$ in Eq.~(\ref{Kc}), the criteria for the groove closure can be rewritten, up to order 2 in $d/h$, by the following inequality  : 
\begin{equation}
\label{Big-lambda}
\Lambda \equiv \frac{8Bd^2}{27 \gamma h^3 \ell_c^2}\left[1+2\frac{d}{h}+0.556 \left( \frac{d}{h}\right)^2\right]^{-1} < 1.
\end{equation}

\subsection{Short-time dynamics.}

Since Eq.~(\ref{eq-ell-adim}) is solved with the initial condition $\bar{\ell_I} = 0$ at $\bar{t}=0$, Eq.~(\ref{eq-beta-adim}) implies that $\beta$ is vanishing at $t=0$ and hence small at short time. Taking the limit $\beta \to 0$ in Eq.~(\ref{eq-ell-adim}) and in the right-hand side of Eq.~(\ref{eq-beta-adim}) and considering $\bar{\ell_I} \ll 1$ in the right-hand side of Eq.~(\ref{eq-ell-adim}), the system of Eqs.~(\ref{model-adim}) becomes
\begin{subequations}
\label{model-adim-short-time}
\begin{align}
\label{eq-ell-adim-short-time}
& 2\, \bar{\ell_I}\, \dot{\bar{\ell_I}} = c(\bar{d}), \quad c(\bar{d})=[\bar{d}^{\, 2}\, G(0,\bar{d})]^{-1},\\
\label{eq-beta-adim-short-time}
& \beta = \left[\frac{1+2\bar{d}}{K\,\bar{d}^{\, 2}}\right] \bar{\ell_I},
\end{align}
\end{subequations}
where, for $\bar{d} \lesssim 1$, 
\begin{equation}
\label{cdbar}
G(0,\bar{d}) = \frac{24 \sqrt{3}\, \bar{d}}{2 \sqrt{3}-\bar{d}} \quad \Rightarrow \quad c(\bar{d}) = \frac{2 \sqrt{3}-\bar{d}}{24 \sqrt{3}\, \bar{d}^{\, 3}}.
\end{equation}


Eq.~(\ref{eq-ell-adim-short-time}) is easily integrated and yields
\begin{equation}
\label{ell-short-time2}
\bar{\ell}_I^{\text{st}}(\bar{t}) = \sqrt{c}\, \bar{t}^{\, 1/2},
\end{equation}
where the superscript st stands for ``short-time''. 
Substituting Eq.~(\ref{ell-short-time2}) in Eq.~(\ref{eq-beta-adim-short-time}), we get
\begin{equation}
\label{beta-short-time}
\beta^{\text{st}}(\bar{t}) = \left[\frac{\sqrt{c}\, (1+2\bar{d})}{K\,\bar{d}^{\, 2}}\right]\,\bar{t}^{\, 1/2},
\end{equation}
where $c$ is given by Eq.~(\ref{cdbar}). 

\begin{figure}[!t]
  \centering
    \includegraphics[width=\columnwidth]{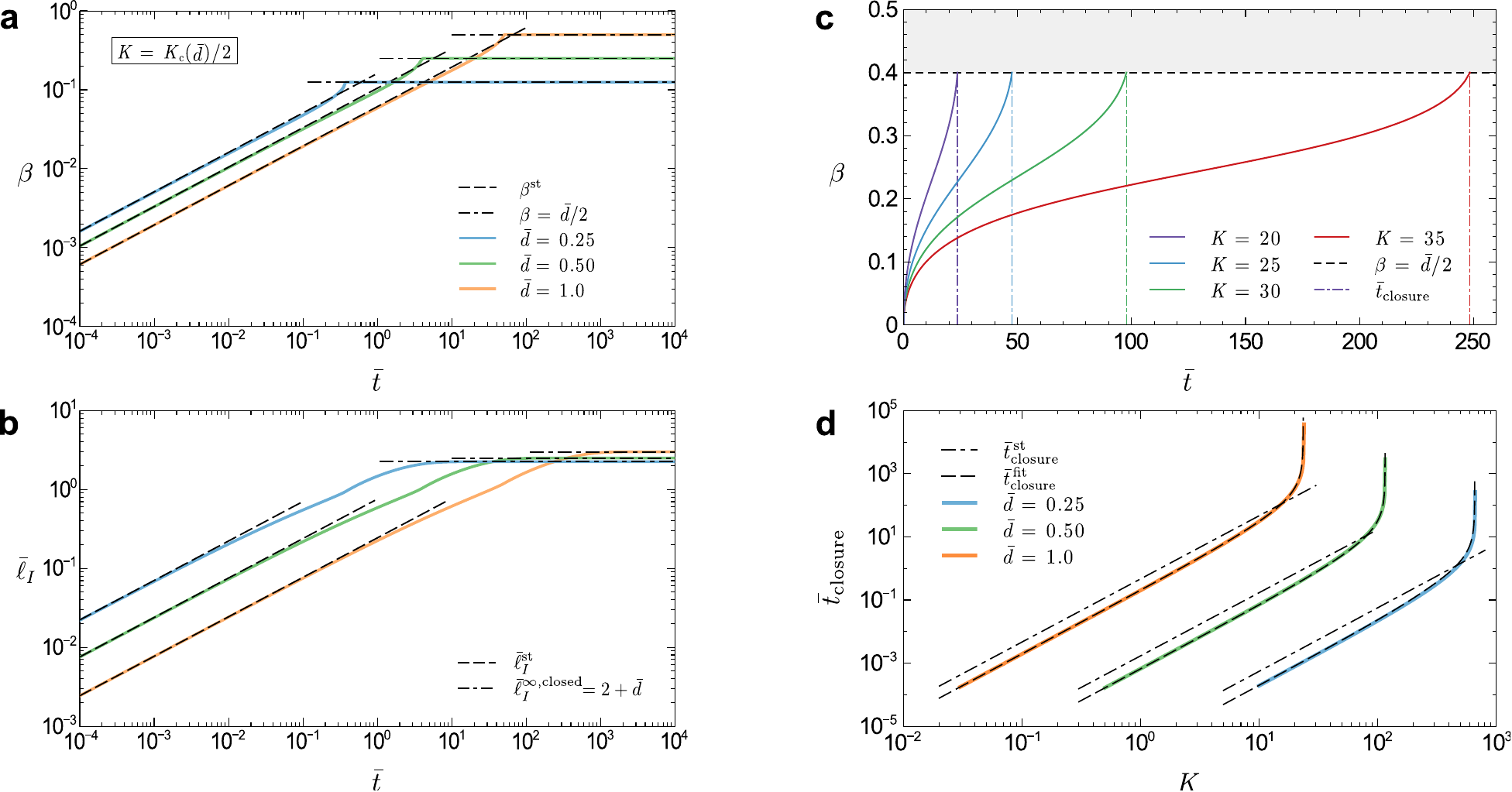}
       \caption{Comparison between the solutions for $\beta$ ({\bf a}) and $\bar{\ell}_I$ ({\bf b}) obtained by solving numerically Eqs.~(\ref{model-adim}), with $K=K_c(\bar{d})/2$ and several values of $\bar{d}$ as indicated, and its asymptotic behaviour at short time ($\beta^{\text{st}}$ and $\bar{\ell}_I^{\text{st}}$, see Eqs.~(\ref{ell-short-time2}) and (\ref{beta-short-time})) and the stationary solution at long time. {\bf c.} Evolution of $\beta$ as a function of time for $\bar{d}=0.8$ and several value of $K<K_c = 38.84$. The closure times at which $\beta=\bar{d}/2$ are shown. {\bf d.} Evolution of the closure time, obtained from Eq.~(\ref{tc-exact}), as a function of $K$ for several values of $\bar{d}$ as indicated. The estimation (\ref{tclosest}) of the closure time obtained from the short-time dynamics is shown with dashed-dotted curves and the corrected expression (\ref{tclosefit}) is displayed with dashed curves.}
       \label{fig_beta_ell}
\end{figure}

Extended Data Figures~\ref{fig_beta_ell}a and b show comparisons between these short time behaviours of $\beta$ and $\bar{\ell}_I$ and the numerical solutions of Eqs.~(\ref{model-adim}), respectively. The dimensionless stiffness is set to $K=K_c(\bar{d})/2$ so that the groove closes and the system reaches the stationary solution $\beta = \bar{d}/2$ and $\bar{\ell}_I=\bar{\ell}_I^{\, \infty,\text{closed}} \to 2 + \bar{d}$ at long time.


\subsection{Time needed for groove closure.}

To determine the time needed for a groove to close, i.e. the time needed for $\beta$ to reach $\bar{d}/2$, Eqs.~(\ref{model-adim}) are first uncoupled to obtain an equation for $\beta$. To do so, $\bar{\ell}_I$ is obtained as a function of $\beta$ from Eq.~(\ref{eq-beta-adim}) and substituted in Eq.~(\ref{eq-ell-adim}):
\begin{subequations}
\label{ODE-beta-all}
\begin{align}
\label{ODE-beta}
&2 K^2 G_1(\beta,\bar{d})\, \beta\, \dot{\beta} = 1 + \beta - K G_2(\beta,\bar{d})\, \beta, \\
\label{eq-G1}
&G_1 = \bar{d}^{\, 2} (\bar{d}-\beta)^3\, \frac{ [\beta (\beta-4 \bar{d}-2)+\bar{d} (1+2 \bar{d})]}{(1+2 \bar{d}-\beta)^3}\, G, \\
\label{eq-G2}
&G_2 = \frac{(\bar{d}-\beta)^2}{1+2 \bar{d}-\beta},
\end{align}
\end{subequations}
where the function $G$ is given by Eqs.~(\ref{Gtilde-def}) and (\ref{G-def}).

Since the function $G$ is strictly positive, the function $G_1$ vanishes only when the expression between square brackets in Eq.~(\ref{eq-G1}) is equal to zero or, equivalently, when $\beta=\beta^*$ with
\begin{equation}
\beta^* = 1+ 2 \bar{d} -\sqrt{1+3\bar{d} + 2 \bar{d}^{\, 2}} = \frac{\bar{d}}{2}+ \frac{\bar{d}^{\, 2}}{8}+\mathcal{O}(\bar{d}^{\, 3}).
\end{equation}
Therefore, as $\beta$ approaches $\beta^*$, which can never be reached because $\beta \le \bar{d}/2$, $G_1$ becomes small and $\dot{\beta}$ must become large to compensate it since the right-hand side of Eq.~(\ref{ODE-beta}) is finite (and strictly positive when $K < K_c$). Note that, because $\beta^*> \bar{d}/2$, $\beta$ can reach continuously its physical extreme value ($\bar{d}/2$ when $K < K_c$) without any discontinuous jump. Indeed, if $\beta^*$ was smaller than $\bar{d}/2$, $\dot{\beta}$ would diverge at $\beta = \beta^*$ and $\beta$ would then jump from $\beta^* < \bar{d}/2$ to $\bar{d}/2$. 

In summary, at short time, $\beta$ grows as $\bar{t}^{\, 1/2}$, as shown by Eq.~(\ref{beta-short-time}), and $\dot{\beta}$ decreases until, if $K < K_c$, $\beta$ approaches $\bar{d}/2$ where $\dot{\beta}$ sharply increases (and $\beta$ as well).

The closure time at which $\beta=\bar{d}/2$ can be obtained exactly. Indeed, Eq.~(\ref{ODE-beta}) is separable and can be integrated side by side yielding
\begin{equation}
2K^2\int_{0}^{\beta} \frac{G_1(\beta',\bar{d})\, \beta'}{ 1 + \beta' - K G_2(\beta',\bar{d})\, \beta'} \, d\beta' = t,
\end{equation}
where the initial condition $\beta(0)=0$ has been considered. Therefore, the closure time is simply obtained by replacing $\beta$ by $\bar{d}/2$ in this relation:
\begin{equation}
\label{tc-exact}
\bar{t}_{\text{closure}} = 2K^2\int_{0}^{\bar{d}/2} \frac{G_1(\beta,\bar{d})\, \beta}{ 1 + \beta - K G_2(\beta,\bar{d})\, \beta} \, d\beta,
\end{equation}
where $G_1$ and $G_2$ are defined by Eqs.~(\ref{eq-G1}) and (\ref{eq-G2}) respectively. Unfortunately, we were unable to compute this integral exactly. Nevertheless, it can be approximated as follow.

As seen in Extended Data Fig.~\ref{fig_beta_ell}a, the intersection between the short-time behavior of $\beta$ given by Eq.~(\ref{beta-short-time}) and its stationary solution $\beta = \bar{d}/2$ gives an estimation the time needed for a groove to close when $K < K_c$. Solving $\beta^{\text{st}}(\bar{t}) = \bar{d}/2$ and using the expression (\ref{cdbar}) of $c$, we have
\begin{equation}
\label{tclosest}
\bar{t}_{\text{closure}}^{\, \text{st}} \sim \left[\frac{6 \sqrt{3}\, \bar{d}^{\, 9}}{\left(2 \sqrt{3}-\bar{d}\right) (2 \bar{d}+1)^2}\right] \,  K^2,
\end{equation}
where the superscript st stands for ``short-time'' since this expression is obtained by extrapolating the short-time behaviour of $\beta$.
However, Extended Data Fig.~\ref{fig_beta_ell}a shows that this expression overestimate the closure time obtained numerically. In addition, the closure time should diverge when $K$ tends to $K_c$ from below because the groove never closes when $K>K_c$. This divergence is not captured by Eq.~(\ref{tclosest}). It is thus expected that Eq.~(\ref{tclosest}) describes well the exact closure time (\ref{tc-exact}) up to a factor and for $K\ll K_c$. However, Eq.~(\ref{tclosest}) can be corrected to better describe the closure time obtained numerically:
\begin{equation}
\label{tclosefit}
\bar{t}_{\text{closure}}^{\, \text{fit}} = \frac{\bar{t}_{\text{closure}}^{\, \text{st}}}{2.4 - 0.2 \ln(\bar{d})/\ln 2}\, \left[1-\frac{K}{K_c}\right]^{-4/5}.
\end{equation}
One factor, which varies slowly with $\bar{d}$, corrects the overestimation of the closure time mentioned above and the second (ad hoc) factor describes the divergence of the closure time near $K=K_c$. Extended Data Figure~\ref{fig_beta_ell}d shows indeed that Eq.~(\ref{tclosest}) overestimates the exact closure time when $K\ll K_c$ and strongly underestimates it when $K \simeq K_c$ since Eq.~(\ref{tclosest}) does not diverge near this value. However, the corrected estimation (\ref{tclosefit}) describes well the exact closure time for any $K$.

\subsection{Second capillary rise dynamics.}

When complete closure of the grooves occurs, a second capillary rise takes place in the newly formed pipe. We assume the number of ribs to be large enough to consider that the pipe section is circular. 
The contact between neighbouring ribs sets the radius of the pipe, $R_{\text{int}}=wh/d$ , but also increases the mechanical resistance of the structure\cite{cappello2023bioinspired} which can be considered as rigid. 

The dynamics of capillary rise in such rigid cylindrical pipes has been already described by Lucas\cite{lucas1918ueber} and Washburn\cite{washburn1921dynamics}. The flow is  assumed unidirectional and we neglect inertial forces that only play a role at very early dynamics. Then, the flow results from the competition between the driving capillary force, due to the pressure gradient induced by the curvature of the liquid-vapor interface, and the resistive gravitational and viscous forces. The flow in the pipe is Poiseuille-like, and the relation between the capillary pressure drop $\Delta p$ and the flow rate $Q = \pi R_{\text{int}}^2 d\ell/dt$ reads: 
\begin{equation}
\Delta p -\rho g \ell_{II} = R_\mu Q,
\label{eq:resistance-hydro}
\end{equation}
where $R_\mu = 8\mu \ell_{II}/(\pi R_{\text{int}}^4)$ is the hydraulic resistance of a cylindrical pipe of radius $R_{\text{int}}$.

The pressure at the top of the liquid column is given by the Laplace law $p(\ell_{II}) = -2\gamma/R + p_{\text{atm}}$ while the pressure at the bottom of the pipe is equal to the atmospheric pressure $p(0) = p_{\text{atm}}$. Hence, the pressure drop is $\Delta p = 2 \gamma /R$.

Using the expressions of the flow rate, $Q$, and of the capillary pressure drop, $\Delta p$, in Eq.~(\ref{eq:resistance-hydro}) the evolution equation for liquid height $\ell_{II}$ in
the tube: 
\begin{equation}
\frac{8\mu}{R_\text{int}^2}\ell_{II}\dot{\ell}_{II} = \frac{2\gamma}{R_{\text{int}}}-\rho g \ell_{II},
\label{eq:dynamics-ell2}
\end{equation}
where $\dot{\ell_{II}} = \text{d}\ell_{II}/\text{d}t$.
For convenience, the evolution equation Eq.~(\ref{eq:dynamics-ell2}) can be made dimensionless by rescaling the variables as follows :

\begin{equation}
\bar{\ell}_{II} = \ell_{II}/\ell_{II}^{\, \infty}, \quad \bar{t} = t/ \tau_{II},
\end{equation}
where 
\begin{equation}
\ell_{II}^{\, \infty} = 2 \frac{\ell_c^2}{R_{\text{int}}}, 
\end{equation}
is the equilibrium height of the second capillary rise (i.e. the Jurin height) obtained by setting $\dot{\ell_{II}}=0$ in Eq.~(\ref{eq:dynamics-ell2}), and $\tau_{II}$ is given by : 
\begin{equation}
\quad \tau_{II} = \frac{16\mu \gamma}{\rho^2 g^2 R^3_{\text{int}}}.
\end{equation}

In dimensionless form, Eq.~(\ref{eq:dynamics-ell2}) becomes
\begin{equation}
\bar{\ell}_{II}\dot{\bar{\ell}}_{II} = 1-\bar{\ell}_{II}.
\end{equation}
This first order nonlinear ODE can be solved numerically using Matlab or Mathematica with the initial condition $\bar{\ell}_{II}(\bar{t}_\text{closure})=0$. For $\bar{t}<\bar{t}_\text{closure}$ no second capilary rise occurs as the the whole structure has not deformed into a tube and the the second capillary rise is set to zero, $\bar{\ell}_{II}(\bar{t}<\bar{t}_\text{closure})=0$.


\subsection{Full immersion of the structure into the liquid bath.}

In this section we propose a model to describe the dynamics of the capillary rise in structures that are plunged in the liquid bath at a velocity $V$ until fully immersed and then withdrawn from this bath with the same velocity.
We compare the cases of the immersion of a rigid structure, that has the same geometry as the tubular-shaped closed flexible structure, and of a flexible grooved sheet. 

Let us first consider the case of a rigid structure. The situation differs from what has been described in the previous sections as, first, the structure being rigid, $\beta$ is constant and equal to $\beta = \bar{d}/2$, and second, while the structure is immersed into or withdrawn from the bath the pressure at the bottom of the structure varies with the length of immersion $L'$.
In the immersion phase $L' = V t$, while in the withdrawing phase $L' = V(t^*-t)$, with $t^* = L/V$ the time at which the structure is fully immersed and withdrawing starts. Hence, the extra pressure drop is $\Delta p_{\text{im}} = \rho g Vt$ during the immersion phase and $p_{\text{im}} = \rho g V(t^*-t)$ in the withdrawing phase.

The dynamics of the capillary rise $\ell_I$ in the pipe of circular sector cross-section is thus described by :
\begin{subequations}
\label{eq:dyn-groove-rigid-imm-all}
\begin{align}
\frac{\mu h}{2d^3} G(\beta=\bar{d}/2, \bar{d}) \ell_I\dot{\ell_I} = &\frac{2\gamma(2+\bar{d})}{d}+ \rho g V t  -\rho g  \ell_I
\label{eq:dyn-groove-rigid-imm}\\ \nonumber &\text{for } 0\leq t<t^*, \, \text{and}\\
\frac{\mu h}{2d^3}  G(\beta=\bar{d}/2, \bar{d}) \ell_I\dot{\ell_I} = &\frac{2\gamma(2+\bar{d})}{d}+ \rho g V (t^*-t)  -\rho g  \ell_I \label{eq:dyn-groove-rigid-with}\\ \nonumber &\text{for }  t^*<t\leq 2L/V.
\end{align}
\end{subequations}
Note that these equations correspond to Eq.~(\ref{eq-ell-temp}), for $\beta = \bar{d}/2$, and hence $A = hd/2$, to which is has been added the aforementioned extra pressure term. By rescaling $\ell_{I}$, $t$ and $V$ in Eqs. (\ref{eq:dyn-groove-rigid-imm-all}) by $\ell_{I}^{\, \infty,\text{open}}$, $\tau_{I}$, and $\ell_{I}^{\, \infty,\text{open}}/\tau_{I}$, respectively, we obtain the dimensionless form of the evolution equation of the capillary rise into the closed grooves:
\begin{subequations}
\begin{align}
G(\beta=\bar{d}/2, \bar{d}) \bar{d}^{\, 2} \bar{\ell}_I\dot{\bar{\ell}}_I = &(2+\bar{d})+ \bar{V} \bar{t}  -\bar{\ell}_I
\label{eq:dyn-groove-rigid-imm-adim}\\ \nonumber &\text{for } 0\leq \bar{t}<\bar{t}^*, \, \text{and}\\
G(\beta=\bar{d}/2, \bar{d}) \bar{d}^{\, 2} \bar{\ell}_I\dot{\bar{\ell}}_I = &(2+\bar{d})+ \bar{V} (\bar{t}^*-\bar{t})  -\bar{\ell}_I \label{eq:dyn-groove-rigid-with-adim}\\ \nonumber &\text{for }  \bar{t}^*<\bar{t}\leq 2\bar{t}^*.
\end{align}
\end{subequations}
These first order nonlinear ODEs are solved numerically using Matlab with the initial condition $\bar{\ell}_I(0) = 0$.

Following the same reasoning, Eq.~(\ref{eq:dynamics-ell2}) describing the capillary rise $\ell_{II}$ in the inner tube may be transformed to take into account the extra pressure term: 
\begin{subequations}
\label{eq:dyn-tube-rigid-imm-all}
\begin{align}
\frac{8\mu}{R_\text{int}^2}\ell_{II}\dot{\ell}_{II} = \frac{2\gamma}{R_{\text{int}}}+\rho g V &t-\rho g \ell_{II}\label{eq:dyn-tube-rigid-imm}
 \\\nonumber & \text{for }\,0\leq t<t^*\,\text{and} \\
\frac{8\mu}{R_\text{int}^2}\ell_{II}\dot{\ell}_{II} = \frac{2\gamma}{R_{\text{int}}}+\rho g V &(t^*-t)-\rho g \ell_{II} \label{eq:dyn-tube-rigid-with}\\ 
\nonumber   & \text{for }\, t^*<t\leq 2L/V.
\end{align}
\end{subequations}
To obtain the dimensionless form of Eqs.~(\ref{eq:dyn-tube-rigid-imm-all}) we rescale $\ell_{II}$ by $\ell_{II}^{\, \infty}$, $t$ by $\tau_{II}$ and $V$ by $\ell^{\infty}_{II}./\tau_{II}$. It leads to 
\begin{subequations}
\begin{align}
\bar{\ell}_{II}\dot{\bar{\ell}}_{II} &= 1+\bar{V} \bar{t}-\bar{\ell}_{II} \quad \text{for }\,0\leq t<t^*\,\text{and}\label{eq:dyn-tube-rigid-imm-adim}\\
\bar{\ell}_{II}\dot{\bar{\ell}}_{II} &= 1+\bar{V} (\bar{t}^*-\bar{t})-\bar{\ell}_{II} \quad \text{for }\, \bar{t}^*<\bar{t}\leq 2\bar{t}^*.
\end{align}
\end{subequations}
A Matlab routine is used to solve these ODEs with the initial condition $\bar{\ell}_{II}(0) = 0$.

For the case of the immersion of the flexible grooved sheet, we consider that during the immersion phase $\ell_{I} = V t$ since while the structure is plunged in the bath the structures remains open (see Supplementary Video 4 and 5) and the liquid fills the immersed portion of the structure from the side of the grooves almost instantaneously. To model the capillary rise dynamics in the withdrawing phase, we assume that, at the transition between the immersion phase and the withdrawing phase (i.e. at $t=t^*$), the structure deforms into a tube, therefore $\ell_{II}$ jumps to the value $\ell_{II}(t=t^*) = L$. Then, while the structure -- that remains in its deformed state -- is withdrawn from the bath (i.e. for $t>t^*$), the captured liquid drains.
The equations that govern the drainage in the grooves and in the newly formed tube are the same as in the rigid case, Eq.~(\ref{eq:dyn-groove-rigid-with}) and (\ref{eq:dyn-tube-rigid-with}), respectively. However, the initial conditions change and are $\ell_{I}(t=t^*) = \ell_{II}(t=t^*)= L$. 

A MatLab routine is used to solve the first order ODEs and obtain the height of the first and second capillary rise at the end of the withdrawing phase, $\ell_{I}(t = 2t^*)$ and $\ell_{II}(t = 2t^*)$.

Examples of capillary rise dynamics during the entire dipping of flexible or rigid structures in the liquid bath are shown in Fig.~\ref{fig3}c of the article and in
Supplementary Videos 4 and 5, and the evolution of the final capillary rise heights as a function of the velocity $V$ is illustrated in Fig.~\ref{fig3}d of the article.

\subsection{Discussion on inertial effects.}

In all our calculations we completely disregarded inertial terms. However, at early stages inertia impacts the flow dynamics and leads to an extra term in the fluid momentum equation. The expression of such a force scales as 
\begin{equation}
F_\text{in} \sim  \rho  A \ell_{I,II} \frac{d^2\ell_{I,II}}{dt^2},
\end{equation}
 with $\ell_{I,II}$ being either the first or the second capillary rise height. The inertial term has to be compared to the viscous force, that, when $d\sim w\sim h \sim R_{\text{int}}$, as in the case in our experiments, scales as 
\begin{equation}
F_\mu \sim  (\mu A^2/d^4) \ell_{I,II} \frac{d\ell_{I,II}}{dt}.
\end{equation}
Balancing the two forces, $F_\text{in}$ and $F_\mu$, we obtain the typical timescale $\tau_\text{in} = \rho d^2/\mu$ below which inertia plays a significant role and above which it can be disregarded compared to viscous forces. Experimentally, we used structure of typical dimension $d\sim 700$~µm, fluid density $\rho = 1000$~kg/m$^3$ and of viscosity equal or larger than $\mu = 10$~mPa s. For such values, $\tau_\text{in} \sim 5 \times 10^{-2}$~s. 

\end{document}